\documentclass[
prb,twocolumn,
amsfonts,amssymb
]{revtex4-2}

\usepackage{natbib}

\usepackage{graphicx}
\usepackage{color}
\usepackage{dcolumn}
\usepackage{bm}
\usepackage{psfrag}
\usepackage{ulem}

\begin{document}
\title{Linear and non-linear current response in disordered d-wave superconductors} 
\author{L. Benfatto}
\affiliation{Dipartimento di Fisica, Universit\`a di 
  Roma ``La Sapienza'', and ISC-CNR, P.$^{le}$ Aldo Moro 5, 00185 Roma, Italy}
\author{C. Castellani}
\affiliation{Dipartimento di Fisica, Universit\`a di 
  Roma ``La Sapienza'', and ISC-CNR, P.$^{le}$ Aldo Moro 5, 00185 Roma, Italy}
\author{G. Seibold}
\affiliation{Institut f\"ur Physik, BTU Cottbus-Senftenberg, D-03013 Cottbus, Germany}
\date{\today}

\begin{abstract}
We present a detailed theoretical investigation of the linear and non-linear optical response in a model system for a disordered $d$-wave superconductor, showing that for both quantities  the gap symmetry considerably changes the paradigm of the optical response based on the conventional $s$-wave case. For what concerns the linear response our findings agree with previous work showing that in strongly-disordered $d$-wave superconductors a  large fraction of uncondensed spectral weight survives below $T_c$, making the optical absorption around the gap-frequency scale almost unchanged with respect to the normal state. Our numerical results are in excellent quantitative agreement with experiments in overdoped cuprates. 
 In the non-linear regime we focus on the third-harmonic generation (THG), finding that, as already established for the $s$-wave case, in general a large THG is triggered by disorder-activated paramagnetic processes. However, in the $d$-wave case the BCS response is monotonously increasing in frequency, loosing any signature of THG enhancement when the THz pump frequency $\omega$ matches the gap maximum $\Delta$, a hallmark of previous experiments in conventional $s$-wave superconductors. Our findings, along with the mild polarization dependence of the response, provides an explanation for recent THG measurements in cuprates, setting the framework for the theoretical understanding of non-linear effects in unconventional cuprates. 

\end{abstract}

\maketitle

\section{Introduction}
Since the discovery of high-critical temperature ($T_c$) superconductivity in cuprates, optical-conductivity measurements
have been central in advancing our understanding of their
unusual electronic properties, including e.g. the symmetry of the superconducting (SC) gap, the opening of a pseudogap above $T_c$, or the transition from a Mott-insulating state to a (non-) Fermi liquid with increasing doping (for a review see e.g. ~\cite{basov_rmp11,tajima_rpp16}). In more recent years, the development of non-equilibrium spectroscopies, made possible by the use of intense light pulses, offered the potential opportunity to disentangling different dynamical processes at play in complex systems via their different relaxation
times~\cite{giannetti_review}. With regard to SC materials, the advent of THz spectroscopy at strong fields turned out to be particularly promising, due to the frequency-matching condition between the light and the typical energy scales  at play in the SC phenomenon. This condition allows one to observe relevant effects already with pump-only experiments, thanks to the ability of the intense light pulses to trigger via non-linear optical processes collective excitations invisible in the linear response.  This is the case e.g. for the long-sought amplitude or Higgs mode of the superconductor, that couples to light only to quadratic order, being charge neutral\cite{shimano_review19}. 

A paradigmatic demonstration of non-linear THz processes is provided by the measurement of enhanced third-harmonic generation (THG) below $T_c$, that has been demonstrated so far in a variety of systems, ranging from conventional NbN \cite{shimano_science14,shimano_prb17,wangNL_NbN_prb22} and MgB$_2$\cite{wang_mgb2_prb21,wang_mgb2_prb22} superconductors, and more recently in unconventional pnictides \cite{shimano_commphys21} and cuprate superconductors
\cite{kaiser_natcomm20,shimano_prb23,averitt_prb23,wangNL_cm22,kaiser_natcomm23,kaiser_cm23}. From the theoretical point of view, the interpretation of these experiments can still rely on a quasi-equilibrium scheme, provided that the optical response is computed beyond linear order. 
Basically, the current density in response to an applied vector potential ${\bf A}(t)$ can be expanded up to third order  as
\begin{equation}
  j_\alpha=\chi_{\alpha\beta}^{(1)}A_\beta + \chi^{(3)}_{\alpha\beta\gamma\delta}
  A_{\beta}A_\gamma A_{\delta}
\end{equation}
where $\chi^{(1)}$ is the linear response which is related to the
optical conductivity, and $\chi^{(3)}$ is the non-linear optical kernel. 
Despite such a considerable simplification with respect to the pure non-equilibrium phenomena, the interpretation of the THG experiments on superconductors stimulated so far considerable theoretical work\cite{aoki_prb15,aoki_prb16,cea_prb16,cea_prb18,silaev_prb19,shimano_prb19,tsuji_prr20,schwarz_prb20,seibold_prb21,eremin_bs_prb21,fiore_prb22,udina_faraday22}.
The reason is that in a $s$-wave superconductor both the direct contribution to $\chi^{(3)}$ of BCS quasiparticles and of the amplitude fluctuations of the order parameter give the largest response to THG when the frequency $\omega$ of the THz pump matches  the value of the SC gap, $\omega=\Delta$, making it difficult to disentangle the two effects in the experimental results. In the case of $s$-wave superconductors it has been so far established that the hierarchy among the two contributions is ruled in a crucial way by disorder effects. Indeed, while for clean single-band\cite{cea_prb16} and multi-band\cite{cea_prb18} $s$-wave superconductors the THG is dominated by diamagnetic-like processes, yielding a predominant response from BCS quasiparticles, disorder makes possible also paramagnetic-like processes which also couple non-linearly the light to the system\cite{silaev_prb19,shimano_prb19,tsuji_prr20,seibold_prb21}. The consequences are twofold: from one side, the strength of the THG is overall enhanced even by weak disorder while still retaining  a rather sharp resonance at $\omega=\Delta$, explaining thus the rather large effects measured in the experiments. From the other side, at relatively strong disorder  the Higgs response can eventually overcome the BCS one, allowing for a preferential channel to drive the Higgs mode by light. In addition, disorder influences also the polarization dependence of the THG signal, i.e. the dependence of the generated non-linear current on the angle that the applied field ${\bf A}$ forms with the main crystallographic axes of the lattice. So far, numerical studies\cite{seibold_prb21,udina_faraday22} accounting exactly for disorder effects on a prototypical square lattice showed that the THG response, that is strongly polarization-dependent in the clean limit\cite{cea_prb16}, becomes rather isotropic already at moderate disorder, in agreement with usual observations in both $s$-wave\cite{shimano_prb17,wangNL_NbN_prb22}  and $d$-wave\cite{kaiser_natcomm20,kaiser_natcomm23,wangNL_cm22} superconductors.

For d-wave superconductors theoretical studies of THG have been restricted until now to the clean limit\cite{schwarz_prb20}, confirming the predominance of the BCS response in this regime. In addition, these studies find that even though in a $d$-wave superconductor a continuum of BCS excitations exists below twice the gap maximum $\Delta$ even at $T=0$, the non-linear kernel $\chi^{(3)}$ preserves in the clean limit a strong resonance at $2\Delta$ and a pronounced polarization dependence. However, both effects seem to be in contrast with the experimental observations in cuprates\cite{kaiser_natcomm20,kaiser_natcomm23,wangNL_cm22,kaiser_cm23}: indeed, in these systems the THG signal has a rather smooth temperature dependence, with no clear signatures of a resonance effect at the temperature where the pumping frequency matches the maximum gap value, and it has a mild polarization dependence. In addition, the THG is found to persist on a wide range of temperatures above $T_c$, calling for the possible contribution of fluctuations effects due to the THz probing frequency\cite{gabriele_natcomm21}. 

In this manuscript we address explicitly the role of disorder on the THG response of a $d$-wave superconductor. 
In general, cuprates are usually expected to be in the relatively clean limit, at least for underdoped and optimally-doped samples, where optical-conductivity measurements in the THz regime\cite{shimano_prb20} suggest a value $1/(2\Delta\tau) \simeq 0.85$, with $\tau$ being the transport scattering rate. Nonetheless, already such a small disorder can significantly enhance the THG response and affect the polarization dependence, as it has been shown\cite{udina_faraday22} in some recent work with the band dispersion of cuprates and $s$-wave order-parameter symmetry.

The situation can be eventually different for strongly overdoped cuprates, as e.g. in  overdoped LSCO
films \cite{armitage_prl19,armitage_prb22}, where already the linear optical response revealed a significant uncondensed fraction of charge carriers and a concomitant 'Drude'-like behavior of $\sigma_1(\omega)$ even below $T_c$.
These data, together with the experimentally found correlation between
$T_c$ and the superfluid density \cite{bozovic_nat16} in the overdoped regime,
have stimulated the idea that disorder, presumably
due to out-of plane dopant ions, is a key player in understanding the
vanishing of $T_c$ in this doping range\cite{hirschfeld_prb18,leedh_npjqm21,hirschfeld_prb23}. 

To address these issues we compute the 
linear and non-linear current response within a disordered lattice model where the $d$-wave SC order is induced via a Heisenberg-like spin-spin interaction. As compared to previous work in the $s$-wave case\cite{seibold_prb21,udina_faraday22}, the challenge here is the necessity to carry out the numerical simulations  with large lattices,
since the low-energy response is dominated by the nodal regions where the SC gap vanishes. This is indeed the origin\cite{lee_prl93} of the so-called universal value $\sigma_0$ of the optical conductivity in the limit $\omega\to 0$. i.e. a value independent on disorder as long as the scattering rate is much smaller than the maximum SC gap $\Delta$ and vertex corrections can be ignored \cite{lee_prl93,lee_prb00}.
In the present paper we perform a systematic analysis of the linear and non-linear current response as a function of lattice size, using the linear response also as a benchmark of the reliability of the numerical results  even at small disorder  down to frequencies well below $\Delta$. Finite-size effects become instead irrelevant when the scattering
rate becomes of the same order than $\Delta$,  so in this limit our calculated response is essentially valid down to $\omega=0$. The results for the optical conductivity and the superfluid stiffness are in very good agreement with experimental data from cuprate superconductors\cite{armitage_prl19,armitage_prb22}. In particular, in the disorder regime $2\Delta\tau \approx 1$ relevant for overdoped cuprates, our calculations support the relevance of out-of plane impurities for the vanishing of $T_c$ and of the superfluid fraction. 

For the THG our calculations show a marked qualitative difference with respect to the $s$-wave case. Indeed, while the general mechanism of a strong overall enhancement of paramagnetic-like THG processes is confirmed, for $d$-wave pairing the response rapidly looses the resonance at $2\Delta$ characteristic of the clean limit, that is instead found to survive for $s$-wave superconductors up to strong disorder\cite{seibold_prb21}. At the same time, disorder washes out the strong orientation dependence due to the diamagnetic-like processes dominating the clean case, in analogy with the $s$-wave case. Both effects, i.e. the absence of a marked frequency resonance of $\chi^{(3)}$ at $2\Delta$ and the smooth polarization dependence, are in good agreement with the recent experimental findings\cite{kaiser_natcomm20,kaiser_natcomm23,wangNL_cm22,kaiser_cm23},  as discussed above.

The paper is organized as follows: In Sec. \ref{model} we discuss the model
and how the disordered ground state solutions are obtained within the
Bogoljubov-de-Gennes approach. Sec. \ref{resultsigma} presents the corresponding
results for the optical conductivity, that are discussed within context of experiments on the superfluid density in overdoped cuprates.
The influence of disorder on the third harmonic response and its dependence on
the polarization of the incoming light is then analyzed in \ref{resultthg} and we conclude our discussion in Sec. \ref{discussion}.
In the Appendix \ref{sec:thg}  we give a detailed derivation on how
we compute the linear and non-linear current, Appendix
\ref{appb} is devoted to an analysis of finite size effects, Appendix
\ref{sec:s0} discusses the evaluation of the universal conductivity for the parameters used in the present paper,  and finally  
in Appendix \ref{sec:tau} we report the estimate of the transport scattering time for the different
disorder levels we used to evaluate the optical response.

\section{Model}\label{model}
To model the $d$-wave SC order emerging in cuprates we consider a tight-binding model on a square lattice, with an interaction part modelled as intersite spin-spin ($\sim J$) interactions together with local on-site disorder (cf. e.g. \cite{trivedi_prb63,ghosal_prb95,ghosal_prb96})

\begin{eqnarray}                                         
  H&=&\sum_{ij\sigma}(t_{ij}-\mu\delta_{ij}) c^\dagger_{i\sigma}c_{j\sigma} +\sum_{i\sigma} V_i n_{i\sigma} \label{eq:model}\\
  &+& J \sum_{\langle ij\rangle} \left\lbrack {\bf S}_i{\bf S}_j -\frac{1}{4}n_i n_j\right\rbrack  \nonumber                                     
\end{eqnarray}                                                                  
where $t_{ij}$ includes the hopping between nearest ($\sim t$) and next-nearest ($\sim t'$) neighbors, $\langle ij\rangle$ denotes the summation over nearest-neighbors only, and ${\bf S}_i$ is the spin operator at lattice
site $R_i$. $V_i$ is
a random variable taken from a flat distribution with $-V_0 \le V_i \le +V_0$.

In order to avoid any interference with competing phases,
we neglect the decoupling of the interaction part with respect
to on-($i=j$)  and intersite ($i\ne j$) charge densities
$\langle c_{i,\sigma}^\dagger c_{j,\sigma}\rangle$.
The mean-field hamiltonian therefore reads
  \begin{eqnarray}
    H^{MF}&=&\sum_{ij\sigma}(t_{ij}-\mu\delta_{ij}) c^\dagger_{i\sigma}c_{j\sigma}
    +\sum_{i\sigma} V_i n_{i\sigma} \label{eq:modelmf} \\
    &+&\frac{1}{2}\sum_{i,\delta}\left\lbrack \Delta_{i,\delta}\left(
    c_{i,\uparrow}^\dagger c_{i+\delta,\downarrow}^\dagger 
    + c_{i+\delta,\uparrow}^\dagger c_{i,\downarrow}^\dagger \right) \right.\nonumber \\
    &+& \left. \Delta^*_{i,\delta}\left(c_{i,\downarrow}c_{i+\delta,\uparrow}+c_{i+\delta,\downarrow}c_{i,\uparrow}\right)\right\rbrack \nonumber \\
    &+&\frac{1}{J}\sum_{i,\delta} |\Delta_{i,\delta}|^2
\end{eqnarray}    
  where
  \begin{equation}
    \Delta_{i,\delta}= -\frac{J}{2}\left\lbrack \langle c_{i,\downarrow}c_{i+\delta,\uparrow}\rangle +\langle c_{i+\delta,\downarrow}c_{i,\uparrow}\rangle \right\rbrack \\
    \end{equation}
    represent a bond SC order parameter with $\delta \equiv \pm x, \pm y$.
  
In analogy with the well-studied $s$-wave case\cite{trivedi_prb01,trivedi_nphys11,seibold_prl12,lemarie_prb13,trivedi_prb20,seibold_prb21,garcia_prb22},
the hamiltonian Eq.\ (\ref{eq:modelmf}) can be diagonalized by means of the Bogoliubov-de-Gennes (BdG) transformation
\begin{displaymath}
c_{i\sigma}=\sum_k\left[u_i(k)\gamma_{k,\sigma}-\sigma v_i^*(k)\gamma_{k,-\sigma
}^\dagger\right]
\end{displaymath}
which yields the eigenvalue equations:
\begin{eqnarray}
  \omega_k u_n(k)&=&\sum_{j} t_{nj} u_j(k)  + [V_n - \mu] u_n(k) \nonumber \\ &+&\sum_\delta \Delta_{n,\delta} v_{n+\delta}(k)                          
 \label{eq1}\\                                                                  
\omega_k v_n(k)&=&-\sum_{j}t_{nj} v_j(k)  - [V_n - \mu] v_n(k) \nonumber \\ &+&\sum_\delta \Delta^*_{n,\delta}  u_{n+\delta}(k)\label{eq2} \,.
\end{eqnarray}

In the following $u_i(k)$ and $v_i(k)$ are taken to be real.
This excludes e.g. ground states with circular currents which have been
studied in models where disorder is implemented via a variable concentration $n_i$ of
scatterers with fixed impurity strength \cite{leedh_npjqm21,andersen_prb22}.
In our system with a Anderson type of disorder the
calculations with complex $u_i(k)$ and $v_i(k)$ did not yield  stable solutions at small and intermediate disorder. For large $V_0/t \sim 1$ we observe instabilities in the ground state (cf. Sec. \ref{discussion}),
which may be due to circular currents and which will be investigated in detail elsewhere.

Starting from an initial random distribution for the
  anomalous expectation values we
diagonalize the system of equations (\ref{eq1},\ref{eq2}),
and iterate up to a given accuracy ($10^{-8}$) for the $\Delta_{i,\delta}$.
In order to check the stability of the solution we also add random
values to the iterated anomalous expectation values, and check if
a subsequent iteration converges to the same previous solution.
We then evaluate the linear and nonlinear currents from a perturbative
expansion which is outlined in Appendix \ref{appa}, and corresponds to an explicit evaluation of the response function to a given order in the applied vector potential.  The consideration of
disordered $d$-wave superconductors requires the investigation of large
systems in order to capture the responses at low energies.
Here we study lattices between $52\times 52$ and  $68\times 68$, which
allows us to elucidate the role of finite-size effects. On the other hand,
due to the large system sizes it is not possible to include collective (SC amplitude, SC phase and density)
excitations on top of the BdG solution, so that the responses reflect
the quasiparticle contributions only.  The collective-mode contribution has been shown to become sizeable in both the linear\cite{seibold_prl12,seibold_prb17} and the non-linear\cite{seibold_prb21} response of strongly-disordered $s$-wave superconductors. Even though preliminary calculations on small systems reveal that this is also
the case in $d$-wave superconductors, we will focus for the moment on the BCS response only, that is presumably the most relevant one in the underodped and optimally-doped regime of cuprates. For this reason, we implement here a different (and faster) procedure for evaluating the response functions with respect to our previous work\cite{seibold_prb21,udina_faraday22}, since instead of solving the time dependent density matrix for an arbitrary vector potential we directly truncate the equations of motion for each frequency at the desired order.

Results are obtained for a charge density $n=0.875$, and all parameters
are measured with respect to the nearest-neighbor hopping $t$.
The exchange coupling is taken as $J/t=1$ and we also include
a next-nearest-neighbor hopping $t'/t=-0.2$, as appropriate for cuprates.
For the homogeneous system one obtains a $d$-wave gap
\begin{equation}\label{eq:dwave}
\Delta_{\bf k}=\frac{\Delta}{2}\left[\cos(k_x)-\cos(k_y)\right]
\end{equation}
with $\Delta/t=0.316$ and thus a maximum optical gap $2\Delta$, 
which decreases upon including disorder.
In cuprates, gap values extracted from angle-resolved photoemission (ARPES)
on Bi2212 (see e.g. \cite{hashi_natphys14})
can reach values $\Delta=30 - 40 meV$ which in our model
would correspond to hopping values $t=100 - 130 meV$.
This is compatible with our value for the exchange coupling $J/t=1$
and even with tight-binding fits to the dispersion and Fermi surface
as obtained by ARPES \cite{marki_prb05}.
However, for other compounds the hopping is estimated to be in the range
$t=200 - 300 meV$ so that our investigations should be considered
as a qualitative but not necessarily a  quantitative prediction of the
linear and non-linear current response in cuprate superconductors.

\section{Linear response and superfluid stiffness}\label{resultsigma}
The optical conductivity for an applied electric field along ${\bf e}_x$
and the current response along the same direction is obtained from




\begin{eqnarray} \sigma_1(\omega)&=&\pi\left\lbrack -\langle t_x\rangle +\chi_1(\omega=0)\right\rbrack \delta(\omega) - \frac{\chi_2(\omega)}{\omega} \label{eq:res}\\
  \sigma_2(\omega)&=& \frac{-\langle t_x\rangle +\chi_1(\omega)}{\omega} \label{eq:ims}
\end{eqnarray}
where
$\chi_{1,2}$ are the real and imaginary part of the current-current correlation function $\chi(\omega)$, respectively. Within our formalism, cf. appendix \ref{sec:thg}, $\chi(\omega)$ is obtained from $\chi(\omega)\equiv j^{(1)}_{para}(\omega)$ and $\langle t_x\rangle\equiv j^{(0)}_{dia}$, where in general $j_{para/dia}^{(n)}$ refers to the $n$-th term in the expansion in powers of the gauge field of the paramagnetic or diamagnetic electronic current. Since the diamagnetic current response is already linear in the gauge field, the $0$-th term is sufficient to compute the linear response.  
We set $\hbar=e=a\equiv 1$ so that the conductivity
in these units is dimensionless and should be multiplied by $e^2/\hbar$
in order to obtain a two-dimensional $\sigma(\omega)$ in SI units.
In order to obtain the three-dimensional $\sigma(\omega)$ for
  layered cuprates one should multiply the dimensionless conductivity times $e^2/(\hbar d)$, with
  $d$ denoting the interlayer spacing.

\begin{figure}[hhh]
\includegraphics[width=8cm,clip=true]{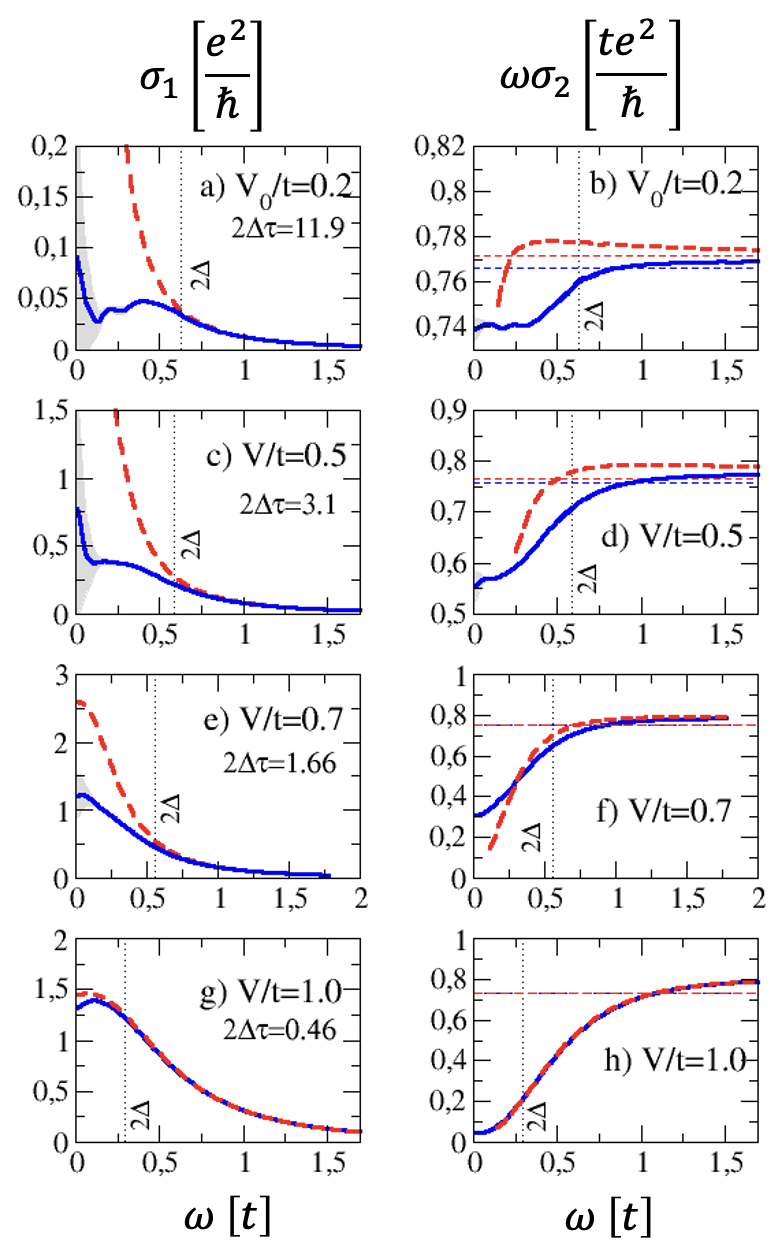}
\caption{Optical conductivity at $T=0$ in the SC state (blue solid line) and in the non-SC state (red dashed line)   for increasing disorder, as  indicated in the panels. Panels (a,c,e,g) show the real part $\sigma_1$, and panels (b,d,f,h) show the product $\omega\sigma_2$ with $\sigma_2$ imaginary part. The limit $\omega \rightarrow 0 $ of $\omega\sigma_2$ identifies the superfluid stiffness, while its high-frequency limit coincides with $-\langle t_x\rangle$, that is weakly disorder dependent. Here we used $J/t=1$, $n=0.875$, $t'/t=-0.2$, and we averaged over different lattice sizes from
  $52\times 52$ to $68\times 68$. The variance with regard to the different
  lattice sizes is indicated by the grey shaded areas. A Lorentzian broadening
(cf. appendix \ref{sec:1st}) $\eta=0.02t$ has been used. The vertical dotted lines indicate the maximum spectral gap $2\Delta$ extracted from the density of states.}     
\label{occup}                                                   
\end{figure}

Fig. \ref{occup} reports real and imaginary part of the optical conductivity
obtained from averages over disorder configurations (($10 - 20$ samples)
and averages over system sizes from $52\times 52 - 68 \times 68$, see also discussion below. Here blue solid lines denote results in the SC state at $T=0$, while red dashed lines denote results at $T=0$ in the absence of SC order. 
On finite lattices and for a given particle number the chemical potential
$\mu$ does not necessarily coincide with
an available quasiparticle energy $\varepsilon_k$, so the minimum
gap for a d-wave superconductor $E_k=\sqrt{(\varepsilon_k-\mu)^2+\Delta^2_k}$ can be different from zero also along the nodal direction.
As shown in Appendix \ref{appb}, for a clean system this minimum gap is an oscillatory function of the system size $N$. Clearly also the low-energy density of states depends on $N$. Nonetheless, this spurious effect can be removed by performing an additional average over systems of different sizes, as shown explicitly in Appendix \ref{appb} in the homogeneous case. 
The grey region in Fig.\ \ref{occup} indicates the spread of the results
due to the evaluation on different lattice sizes. For disorder values up to $V_0/t=0.5$
it can be seen that finite-size effects become relevant in the real part of the optical conductivity for frequencies
below $\omega\sim 0.25t$, whereas they have only minor
influence on $\sigma_2(\omega)$. 
In the strong disorder limit $V/t=1.0$
finite-size effects are no more relevant, so that we can consider the results reliable down to $\omega=0$ in this case.

To better quantify the disorder strength we estimate the dimensionless
quantity $2 \Delta \tau$, where $\Delta$ is the gap maximum obtained for each disorder level from the maximum in the density of states and $\tau$ is the normal-state transport scattering time. To this aim we extract $\tau$ from an analysis of the 
normal-state conductivity, as outlined in Appendix \ref{sec:tau}. In general, by increasing disorder (i.e. by decreasing $2\Delta\tau$), one observes an enhancement of $\sigma_1(\omega)$ in the frequency range below $2\Delta$, as due to the increase of the paramagnetic
current response ($\chi_2(\omega)$ in Eq.\ (\ref{eq:res})), see Fig.\ \ref{occup}a,c,e,g, similarly to previous theoretical studies
\cite{hirschfeld_prb18,leedh_npjqm21,hirschfeld_prb23}. 
For a $d$-wave superconductor it has been shown \cite{lee_prl93} that
the conductivity approaches a so-called universal value $\sigma_0 = \lim_{\omega\to 0} Re\sigma(\omega, T=0)\approx e^2 N_\sigma v_F^2 \hbar/(\pi \Delta)$ for weak disorder, irrespective of the disorder model. Here $N_\sigma$ is the DOS per spin at the Fermi level in the normal state, $v_F$ is the Fermi velocity and the order parameter has the $d$-wave angular dependence $\Delta({\bf k}_F)=\Delta\cos(2\phi)$ on the Fermi surface in the clean case.  In appendix \ref{sec:s0} we evaluate $\sigma_0$ for
the parameters of our model showing that $\sigma_0=2.38 [e^2/\hbar]$ in the weak-disorder limit. On the other hand, as disorder increases $\sigma_0$ is no more universal, and its value depends on the disorder realization. We used the approach of Ref.\ \cite{graf_prb96} to estimate $\sigma_0=0.6 [e^2/\hbar]$ in the strong-disorder unitary limit for our parameter values. These two limiting analytical values can be compared with the numerical estimates of Fig.\ \ref{occup}. 
As mentioned above, for weak to moderate disorder
our values of $\sigma_1(\omega\to 0)$ are subject to
errors due to finite-size effects. Nonetheless, at $V/t=0.5 - 0.7$ the numerical $\sigma_0$ compares well with its expected thermodynamic value within the uncertainty of the
  calculation (denoted by the grey area in Fig. \ref{occup}a,b). In addition, our calculations indicate a low-energy increase of  $\sigma_1(\omega)$, i.e. the tendency for approaching the universal value $\sigma_0$. 
  At strong disorder, where our results for $\omega\to 0$ become reliable, we obtain a somehow larger universal conductivity than what estimated from the perturbative
scheme of Ref. \cite{graf_prb96}, which is expected to fail in this strong-disorder limit. Our calculations also 
indicate a slight downturn upon approaching $\omega \to 0$  which in the
strongly disordered regime \cite{hirschfeld_prl02} develops towards a peak
at $\omega \sim 1/\tau$. In the regime  $2\Delta\tau < 1$ we also observed  the formation of SC islands,
similar to the case of strongly disordered s-wave SC's \cite{trivedi_prb01,trivedi_nphys11,seibold_prl12,lemarie_prb13,trivedi_prb20,garcia_prb22}.

Panels (b,d,f,h) of Fig. \ref{occup} report
$\omega\sigma_2(\omega)=-\langle t_x\rangle + \chi_1(\omega)$ for
the various disorder levels. At large frequencies (larger than the bandwidth) $\chi_1(\omega)\sim 1/\omega$, so that $-\langle t_x\rangle$  identifies  the large-frequency behavior of $\omega\sigma_2(\omega)$. In addition, since $-\langle t_x\rangle$  has very small changes when going from the normal to the SC state, the asymptotic values are pretty much similar in the two cases, as one can see by comparing solid blue and dashed red lines. In contrast, in the limit $\omega \to 0$ this quantity
corresponds to the superfluid stiffness $D_s \equiv \lim_{\omega\rightarrow 0}\omega\sigma_2(\omega)$, which is finite in the SC case but vanishes in the metallic state. In general the stiffness is reduced with increasing
disorder from its clean value $-\langle t_x\rangle$  due to the paramagnetic current response $\chi_1(\omega=0)$. At strong disorder the superfluid stiffness is almost completely suppressed, see panel (h), consistently with the observation of a large fraction of uncondensed spectral weight in $\sigma_1(\omega)$, see panel (g). 

Our results can be compared with experiments on overdoped LSCO films
\cite{armitage_prl19},  which are in the range
$0.31 \lesssim 2\Delta\tau \lesssim 2.45$, 
and overdoped LSCO films, which have been additionally exposed to ion
irradiation \cite{armitage_prb22} leading to even smaller values
of $2\Delta\tau$.
These measurements of $\sigma_1(\omega)$ have revealed that a large fraction
of the carriers remains uncondensed in a wide Drude-type feature at
low temperatures which resembles the $\sigma_1(\omega)$ in panels
(e.g.) of Fig.\ \ref{occup}. A similar residual $\sigma_1(\omega\to 0)$
at low temperatures
has also been observed in Bi2212 thin films.~\cite{corson_prl00}. 

\begin{figure}[hhh]
\includegraphics[width=8cm,clip=true]{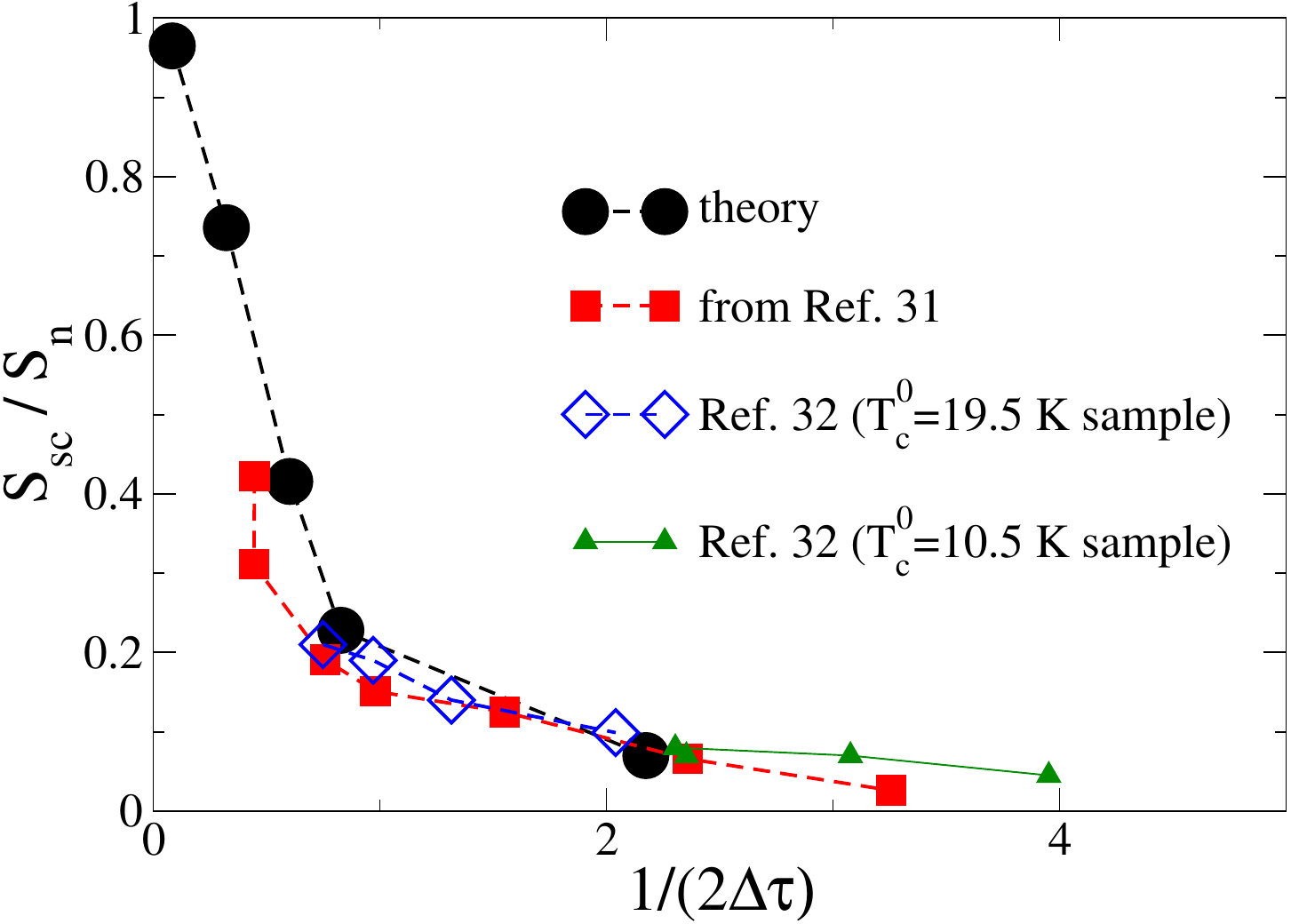}
\caption{Disorder dependence of the ratio $S_{sc}/S_n$ between the superfluid $\delta$-like peak in $\sigma_1(\omega)$ and the total optical spectral weight $S_n$ in the SC state. Our numerical estimates (black dots) are compared with the experimental results from Refs. \cite{armitage_prl19,armitage_prb22}.}     
\label{figarmcmp}                                                   
\end{figure}

To better quantify this effect we follow the same procedure adopted in the experiments\cite{armitage_prl19,armitage_prb22}, by reporting the SC fraction $S_{sc}$, found 
in the SC state as a $\delta$-like peak in $\sigma_1$  at
$\omega=0$, of the total normal-state optical spectral weight $S_n$.  By the so-called  f-sum rule  $S_n=\int_0^\infty d\omega \sigma_1(\omega)=-\pi/2 \langle t_x\rangle$, while 
$S_{sc}$ can be derived from Eqs. (\ref{eq:res})-(\ref{eq:ims}), as given by
\begin{equation}\label{eq:ssc}
  S_{sc}=\frac{\pi}{2}\left\lbrack \langle t_\alpha\rangle +\chi_1({0})\right\rbrack
  =\frac{\pi}{2}\lim_{\omega\to 0} \omega\sigma_2(\omega)\,.
\end{equation}
The ratio $S_{sc}/S_{sn}$
is plotted as a function of $1/(2\Delta\tau)$ in Fig. \ref{figarmcmp}, together
with the corresponding data from Refs. \cite{armitage_prl19,armitage_prb22}. Given the uncertainty in determining the various quantities (e.g. in Ref.\ \cite{armitage_prb22} the
SC gap has been estimated by means of the weak-coupling expression\cite{maki_prb94} for the ratio $\Delta/T_c$) the agreement is rather remarkable.
One could speculate that our model of Anderson-type impurities, where
at each lattice site the local potential is taken from a random
distribution, is more appropriate for the irradiated samples where
the ions affect the electronic structure of the CuO$_2$ planes on a
local scale.
On the other hand, the disorder from the 'unirradiated' data \cite{armitage_prl19}
is presumably mainly due to out-of plane dopant ions which
impose a potential with a finite range on the in-plane
charge carriers\cite{scalapino_prb04}. In view of these considerations it
is interesting that the stiffness obtained in Refs. \cite{armitage_prl19,armitage_prb22} follows nonetheless a similar disorder dependence, which is well captured by our model.

The present calculations of the linear optical response do not include fluctuations effects, that have been shown to be relevant in the strongly-disordered regime for $s$-wave superconductors\cite{seibold_prl12,seibold_prb17,seibold_prb21}. 
As it is well known\cite{schrieffer}, phase fluctuations are crucial to restore the gauge invariance violated by the BCS approximation of the response functions. However, in the clean limit they do not affect the value of the superfluid stiffness, since their correction is purely longitudinal. In the disordered case the longitudinal and transverse response become mixed, and a full gauge-invariant calculation of the stiffness\cite{seibold_prl12} in the disordered $s$-wave case has shown that phase fluctuations lead to a further suppression of $D_s$ as compared to the BCS suppression, along with a pile up of additional finite-frequency optical absorption below the optical gap\cite{seibold_prb17}, analogous to the one shown in Fig.\ \ref{occup}. While  a full computation of $\sigma(\omega)$ adding fluctuations effects is numerically challenging for the  large lattice sizes used so far, we  can nonetheless analyze the static-limit of the fluctuations-induced suppression of $D_s$, following an approach analogous to the one employed in Ref. \cite{seibold_prl12}. We thus add a small ${\cal O}(10^{-3}$) vector potential in the hamiltonian Eq. (\ref{eq:model}) via the Peierls substitution and we compute $D_s$ by taking the ratio between the total current and the vector potential. To this aim, in order to guarantee the  conservation of the total (dia- plus paramagnetic) current at each node, one has to introduce a complex value of the local order parameter. Indeed, one finds 
that not only the local amplitudes $|\Delta_{i,\delta}|$ are affected, but more importantly the local phases are needed to guarantee locally the continuity equation,  that is instead violated by the (non-gauge-invariant) BCS approximation, corresponding to the evaluation in the absence of local-phase relaxation, see Ref.\ \cite{seibold_prl12}.

\begin{figure}[hhh]
\includegraphics[width=8cm,clip=true]{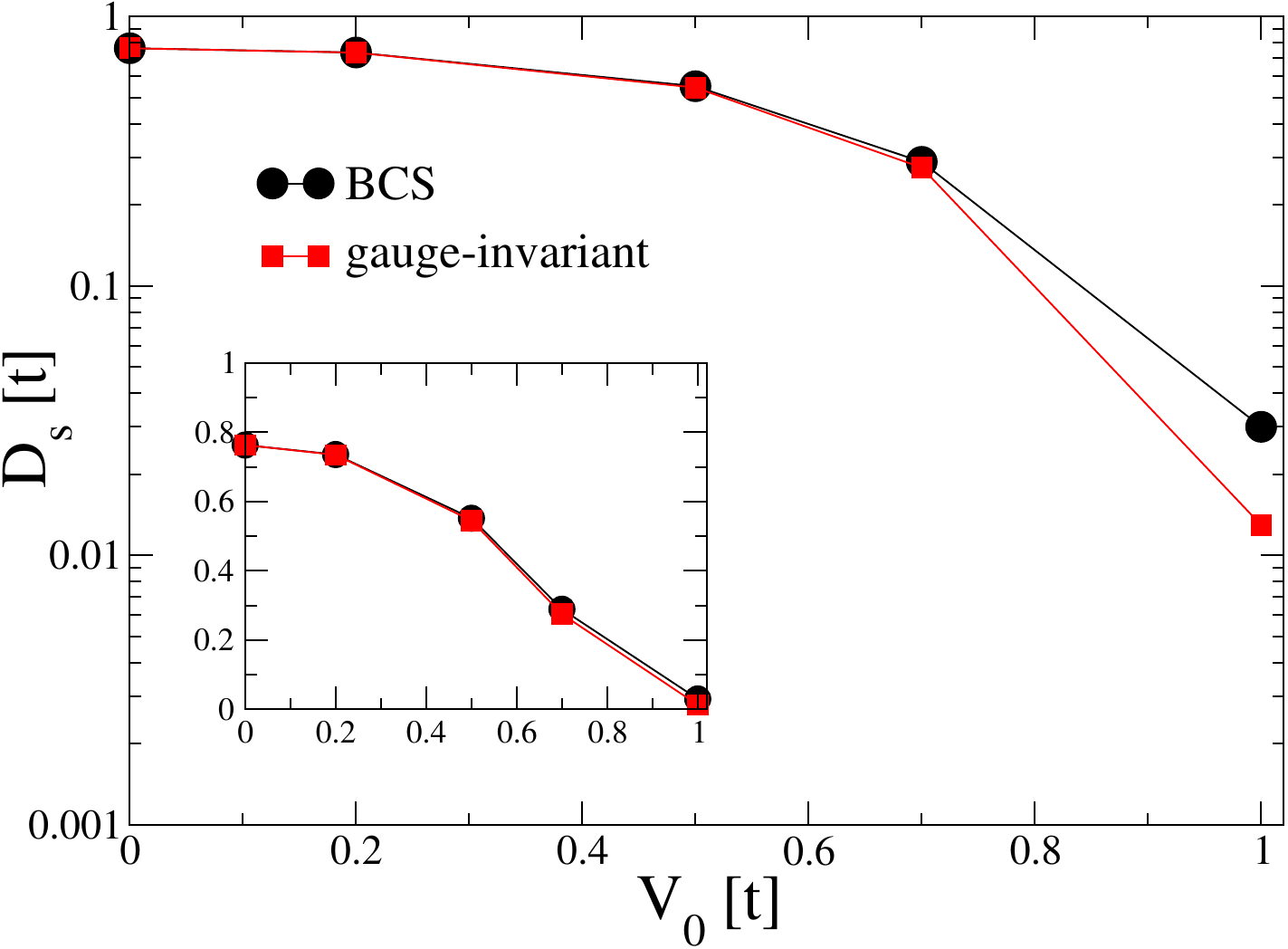}
\caption{Superfluid stiffness $S_{sc}$ as a function of disorder
  from Eq. (\ref{eq:ssc})
  evaluated in the BCS (black) and gauge-invariant (red) approach.}
\label{figbcsrpa}                                                   
\end{figure}

Fig. \ref{figbcsrpa} compares the stiffness, obtained via this gauge-invariant
approach, with the corresponding BCS result on a logarithmic (main panel) and
linear (inset) scale.
We find that for the present parameters fluctuation corrections up to $V_0/t=0.7$ are small ($\sim 5\%$ at $V_0/t=0.7$) whereas at $V_0/t=1.0$ the BCS stiffness
overestimates the gauge-invariant result by a factor $\sim 2$.

We point out that a reduction of $D_s$ due to fluctuation  effects has a significantly stronger impact on the optical conductivity
for the isotropic $s$-wave case rather than for the $d$-wave symmetry. Indeed, the f-sum rule 
  dictates that the decrease in $D_s$, entering the $S_n$ spectral weight of  Eq.\ (\ref{eq:ssc}), has to be compensated by an increase in $\sigma_1(\omega)$. Since collective excitations mostly contribute below $2\Delta$, in a fully-gapped $s$-wave superconductor with a hard gap at $2\Delta$ the effects is a visible sub-gap absorption, as discussed in (\cite{seibold_prb17}). On the other hand in a $d$-wave superconductor quasiparticle excitations already contribute considerable to absorption below $2\Delta$, so that the collective-mode contribution only results in a minor redistribution
  of the finite spectral weight at low energies.

It should be noted that at strong disorder $V_0/t=1.0$ and upon including fluctuations our calculations  reveal
a {\it negative} stiffness for a significant number of disorder realizations. Such samples have been then excluded from the average procedure leading to the results shown in Fig. \ref{figbcsrpa}.
A value $D_s<0$ in principle means that the considered solution does {\it not}
correspond to the true ground state since the free energy as a function
of the vector potential would have a negative curvature. 
In fact, for smaller systems, where an RPA analysis is possible,
  samples with a negative stiffness also lack a well-defined
  zero energy Goldstone mode. The latter instead is replaced by a
  strongly overdamped feature  indicating
that the iteration of the BdG equations has converged to an unstable
solution. In principle, the instability could also arise from the
fact that we have restricted to real valued order parameters which
excludes solutions such as circular currents.~\cite{leedh_npjqm21,andersen_prb22}
Although preliminar investigations have not provided any evidence
for the existence of such currents within our model, we cannot
exclude that there may exist solutions with lower energy which
are made from complex BdG amplitudes.



\section{Third harmonic generation}\label{resultthg}
\begin{figure}[hhh]
\includegraphics[width=8cm,clip=true]{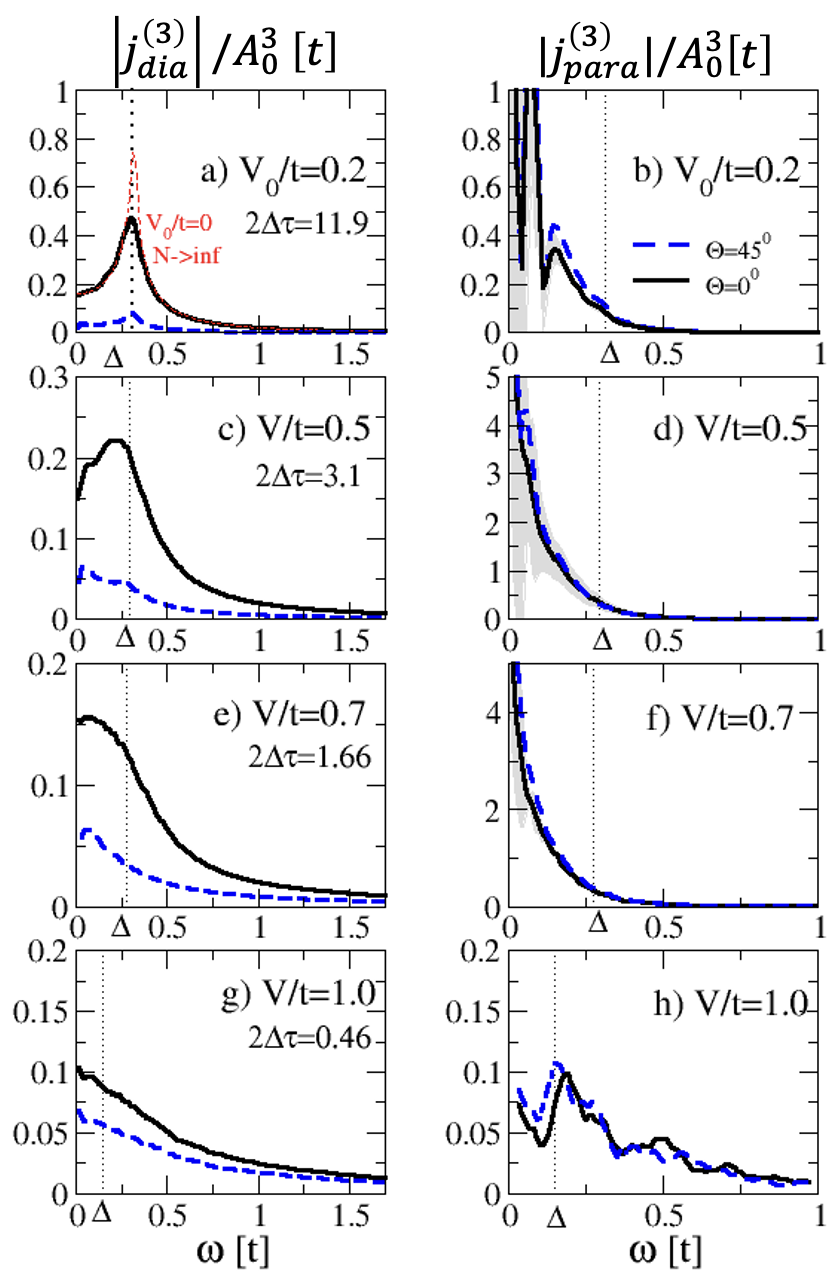}
\caption{Diamagnetic (left column) and paramagnetic (right column) contributions to the third-harmonic current at various disorder levels
  obtained from averages over different lattice sizes from
  $52\times 52$ to $68\times 68$.
The variance with regard to the different
  lattice sizes is indicated by the grey shaded areas.
  Solid black and blue dashed lines refer to the response for a field applied along the $[10]$ and $[11]$ direction, respectively. In panel (a) we also report for comparison the analytical results (\ref{diaclean}) of the $[10]$ clean case, obtained for an infinite-size system. Parameter values are the same as in Fig.\ \ref{occup}. }     
\label{3rdcup}                                                   
\end{figure}

As it has been widely discussed in previous work both for clean\cite{aoki_prb15,cea_prb16,cea_prb18,schwarz_prb20}
 and disordered \cite{silaev_prb19,shimano_prb19,tsuji_prr20,seibold_prb21,udina_faraday22,fiore_prb22} superconductors, the THG can be computed within a diagrammatic approach or within a density-matrix formalism by computing all third-order processes arsing from paramagnetic-like or diamagnetic-like  coupling terms between the gauge field $A$ and the fermions, the former being linear and the latter quadratic in $A$. In full analogy,  as detailed in appendix \ref{sec:thg}, we decompose the third-harmonic current 
in a diamagnetic and paramagnetic contribution $j_{dia/para}(3\omega)$, both computed at finite gauge field and retaining up to third-order in $A$. Since we are interested on the THG measured experimentally with a multicycle pump field,  we study the
the non-linear response to a harmonic vector potential $A(t)=A_0\cos(\omega t)$. In Fig. \ref{3rdcup} we report
the magnitude of both responses $|j_{dia/para}(3\omega)/A^3_0|$ for
the various disorder levels.  We found that these responses are
only finite, within our numerical accuracy, in the SC state. As for the optical conductivity, we have analyzed
finite-size effects which are mainly relevant for the third-harmonic
paramagnetic response in the limit of weak to moderate disorder and low
frequencies (cf. panel b, where the variance is indicated by the grey region).

In the clean limit the third-harmonic paramagnetic current in response to a
monochromatic field at $\omega$ vanishes\cite{cea_prb16,cea_prb18}, and THG is only controlled by the diamagnetic response,
\begin{equation}
\label{diaclean}
  j^{(3)}_{dia,clean}(3\omega)/A_0^3=\frac{1}{N}\sum_k \frac{\Delta_k^2}{E_k}
  \frac{4(t^x_k)^2}{4\omega^2-4 E_k^2} \,,
\end{equation}
whose value  in the limit $N\rightarrow \infty$  is also plotted for comparison with a red dashed line in Fig.\ \ref{3rdcup}.
The $j^{(3)}_{dia,clean}$ corresponds to a density-like correlation function with a diamagnetic vertex\cite{cea_prb16}.
We consider the response along the $[10]$-direction $(\equiv x)$  so that
the vertex is determined by the derivatives of the kinetic energy [i.e. the
Fourier transform of the hopping term in Eq. (\ref{eq:model})]
along the $x$-direction 
$t^x_k=\partial^2\varepsilon_k/\partial k_x^2$ \cite{cea_prb16,schwarz_prb20}.

In the $d$-wave case the logarithmic singularity of $j^{(3)}_{dia,clean}$ at $\omega=\Delta$ in Eq.\ (\ref{diaclean}) is responsible for the strong enhancement of the THG when $\omega=\Delta$, in full analogy with the result of the clean $s$-wave case\cite{cea_prb16}. A strong enhancement is still retained by the diamagnetic response at weak disorder, even though, similarly to the isotropic $s$-wave case\cite{seibold_prb21}, it broadens and decreases with increasing disorder.
Also in analogy with the isotropic $s$-wave case the diamagnetic
current displays a strong orientational dependence. In Fig.\ \ref{3rdcup}
the blue dashed (black solid) line corresponds to $j_{dia}$ along
the $[11]$- ($[10]$-) direction with the field applied along the
same orientation. The suppression of the response along the
$[11]$-direction is most pronounced for weak disorder:  for
$2\Delta\tau=11.9$ one finds a factor of $\approx 6$ between the currents
along $[10]$- and $[11]$-direction. In case of $2\Delta\tau=0.46$ this
value is reduced to $\approx 1.5$.

The paramagnetic current (Fig. \ref{3rdcup}b,d,f,h), absent in the clean case for the same mechanism valid for $s$-wave superconductors\cite{cea_prb16,schwarz_prb20}, becomes finite as soon as even a weak disorder is included. However, in sharp contrast with the disordered $s$-wave case\cite{silaev_prb19,shimano_prb19,tsuji_prr20,seibold_prb21}, it does not retain any resonant behavior around the gap maximum $\omega=\Delta$, but it shows instead a finite value below
$\omega \approx \Delta$, with a continuous increase toward lower
energies. Once more, due to finite-size effects we cannot explore the limit $\omega\to 0$
at weak to moderate disorder. However, the general trend of our results show an increase
of the paramagnetic response with disorder up to $2\Delta\tau \approx 1$,
which then is followed by a decrease in the limit of strong disorder.
In contrast to the diamagnetic current the orientational dependence
of $j_{para}$ is weak at any disorder level, with almost no dependence within the range of the numerical error. This results is also consistent with previous work for the same band structure and $s$-wave pairing\cite{udina_faraday22}, suggesting that the pairing symmetry plays a minor role in determining the polarization dependence of the THG response.

The dramatic increase of the response at low energies together with the absence of a clear resonance in the paramagnetic response represent a marked qualitative difference with respect to the $s$-wave case, where instead the paramagnetic current follows
the diamagnetic response and is enhanced around $\omega=\Delta$\cite{silaev_prb19,seibold_prb21}.  This theoretical result is in agreement with the experimental observation on e.g. conventional $s$-wave superconductors as
NbN \cite{shimano_science14} and MgB$_2$\cite{wang_mgb2_prb21,wang_mgb2_prb22}, where the presence of a frequency maximum in the non-linear kernel $\chi^{(3)}$ can be traced back to an enhancement of the THG measurements at the temperature where the pump frequency $\omega$ matches the SC gap $\Delta(T)$. Interestingly, such a maximum has not been reported so far in most measurements on $d$-wave cuprate superconductors\cite{kaiser_natcomm20,shimano_prb23,wangNL_cm22,kaiser_natcomm23,kaiser_cm23}. One possible interpretation for the lack of a temperature resonance is that in cuprates the SC gap is significantly larger than in conventional superconductors ($\Delta\sim 30-40$ meV)~\cite{hashi_natphys14}. Thus, multi-cycle pulses with central frequency ranging from $\omega=0.5$ to $\omega=0.7$ THz, as the ones used so far, allows one to fulfil the resonance condition only very near to $T_c$, where the  SC response is suppressed. On the other hand, the present calculations showing a rather featureless paramagnetic contribution can explain why for $d$-wave superconductors the non-linear response looses a clear trace of the SC gap value, in contrast to what is found for $s$-wave superconductors. For example the experimental data from Ref. \onlinecite{shimano_prl18} indicate a disorder level $2\Delta\tau \approx 1.2$ where our results (cf. Fig. \ref{3rdcup}, panel f) show a continuosly increasing paramagnetic response towards
lower energy dominating over the diamagnetic one. Translated into a
temperature dependent response via $\Delta(T)$, one would then expect a
continuously decreasing THG which vanishes close to T$_c$, which is in
agreement with the THG data of Ref.\ \cite{kaiser_natcomm20,kaiser_natcomm23,wangNL_cm22}.

\begin{figure}[hhh]
\includegraphics[width=8cm,clip=true]{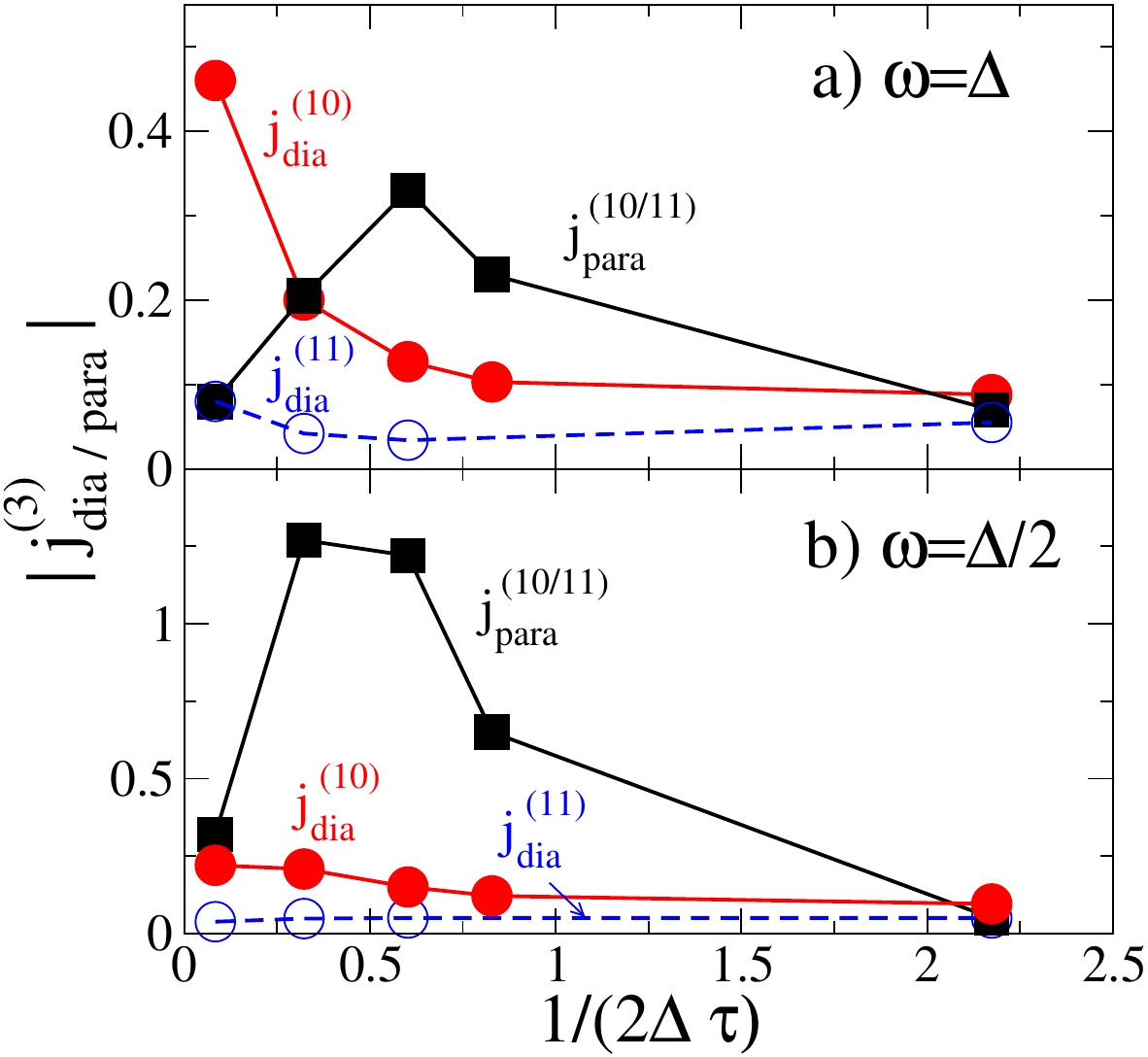}
\caption{Disorder dependence of the diamagnetic (circles) and paramagnetic (squares) response at two fixed frequencies
  $\omega=\Delta_0$ (top) and $\omega=\Delta_0/2$ (bottom). For the diamagnetic response we report the non-linear current in the field direction for a field applied along $(11)$ (blue) and
  $(10)$ (red) direction. For the paramagnetic current the value is the same within the numerical accuracy. }     
\label{cmpiso}                                                   
\end{figure}  

Our theoretical findings are also in excellent agreement with the mild polarization dependence of the THG signal measured experimentally in cuprate superconductors\cite{kaiser_natcomm20,shimano_prb23,wangNL_cm22,kaiser_natcomm23}. Previous work \cite{udina_faraday22} already showed that the inclusion of a next-nearest neighbor hopping $t'$ into the band structure already enhances the isotropy even for $s$-wave pairing. Here the effect is even stronger, and from Fig. \ref{3rdcup} it is evident that for a $d$-wave superconductor an orientational
dependence is only expected for the diamagnetic response. However,
at small frequencies and in the presence of disorder the diamagnetic response is completely
exceeded by the paramagnetic current, which is almost isotropic.
Fig. \ref{cmpiso} summarizes our results showing the non-linear current intensity at two frequency values $\Omega=\Delta$
(panel a) and $\Omega=\Delta/2$ (panel b). Whereas in the former case
dia- and paramagnetic contributions to the THG are comparable, also in strong
contrast to the $s$-wave case \cite{seibold_prb21}, it is apparent that
for smaller frequencies and $1/(2\Delta\tau) \approx 1/2$ the orientational-dependent 
diamagnetic current gives only a minor contribution to the
response. Since for cuprates one is well below the gap maximum for a wide temperature range, our results are pretty much consistent with an isotropic THG signal increasing monotonically below $T_c$. 
Interestingly, despite the lack of a clear signature of the gap maximum in the THG intensity as a function of frequency, preliminary investigations of the phase of the THG response suggest that this quantity could still be sensitive to the crossing at the $\omega=\Delta$ condition. The investigation of this issue will be left for future work. 

\section{Conclusions}\label{discussion}
In the present manuscript we performed a detailed analysis of the linear and non-linear optical
response for a model system of $d$-wave superconductors, intended to reproduce the main features of
cuprate superconductors in the SC state. Since strong correlation effects have not been explicitely included, our analysis is more appropriate for optimally and overdoped samples, even though general aspects, concerning symmetry and disorder also apply to underdoped systems.

The linear response, that has been widely studied in the
past with various analytical and numerical approaches\cite{hirschfeld_prb18,leedh_npjqm21,hirschfeld_prb23}, provides a benchmark to test the accuracy
of our results. Indeed, for a $d$-wave superconductor the vanishing of the gap along the nodal
directions, that is the origin\cite{lee_prl93} of the so-called universal value $\sigma_0$ of the
optical conductivity in the limit $\omega\to 0$,  makes it challenging to provide a reliable
estimate of the zero-frequency limit at small disorder. Nonetheless, we show that our approach,
based on the numerical solution of the self-consistent BdG equations and on a perturbative
expansion in the applied gauge field, allows us to reproduce with good accuracy the expected value
of $\sigma_0$, and the persistence of a large uncondensed fraction of optical spectral weight at
larger disorder, recently reported in overdoped samples\cite{armitage_prl19,armitage_prb22}. 

For what concerns the non-linear optical response we focused on the third-harmonic current generated
in response to a monochromatic gauge field, that has been recently experimentally studied by
different groups in various families of cuprates\cite{kaiser_natcomm20,kaiser_natcomm23,wangNL_cm22,kaiser_cm23}. In analogy with previous work for $s$-wave superconductors \cite{seibold_prb21},
we show that even for small disorder the THG response is strongly enhanced, thanks to the
emergence of a non-linear coupling to light mediated by paramagnetic-like processes, absent in
the clean case. However, in contrast to the $s$-wave case, the paramagnetic response, which
rapidly dominates over the diamagnetic one already for relatively small disorder as appropriate
for cuprates, completely looses any resonance in frequency at the scale set
by the SC gap maximum
$\Delta$. When the system is probed at a fixed pump frequency with varying temperature, this
result would imply a lack of any THG resonance in temperature when cooling below $T_c$. Our
finding, along with the marked isotropy of the THG found in our calculations, bares a strong
resemblance with the experimental data in cuprates, where no clear resonance in temperature of
the THG has been reported\cite{kaiser_natcomm20,kaiser_natcomm23,wangNL_cm22,kaiser_cm23}. 

So far, we only computed the BCS response due to quasiparticle excitations. Indeed, the large system sizes needed to correctly deal with the zero-frequency limit makes it numerically challenging computing also the collective-mode response, as done for the $s$-wave case\cite{seibold_prb21}. It then remains a main open question regarding the fate of the Higgs mode, which in the strongly-disordered $s$-wave case has been found to give a significant (but not predominant at intermediate disorder) contribution to the THG. So far, the analogy with the $s$-wave case lead many authors to interpret also the THG data in cuprates as a signature of the Higgs mode
\cite{kaiser_natcomm20,shimano_prb23,wangNL_cm22,kaiser_natcomm23}, using as main argument the isotropy of the signal as a function of the polarization dependence of the pump. However, the present calculations show clearly, as already anticipated for the same band structure and $s$-wave pairing\cite{udina_faraday22}, that for disorder values consistent with the available experimental results the paramagnetic response due to BCS quasiparticles correctly reproduces the measured isotropy of the THG. In addition, as discussed above, the lack of a resonance in the $d$-wave paramagnetic response could also explain the lack of a temperature maximum in the THG measured so far, showing that for the relatively clean underdoped and optimally-doped samples the BCS response could very well explain the experiments. For what concerns the temperature dependence, a remarkable exception is provided by a recent measurements in an optimally-doped YBCO sample\cite{wangNL_cm22}, where a broad maximum in THG is actually reported near $T_c$. Interestingly, for the same sample the time-trace of the non-linear current also shows a marked beating effect, that seems to be ascribed to an interference effect between two well-defined sharp modes, in analogy with what observed e.g. in NbSe$_2$\cite{chu_cm22}, where the Higgs mode coexists with a CDW amplitude mode.  A possible interpretation of these results could be that along with a featureless BCS paramagnetic response, also a subleading resonant Higgs contribution arises, that becomes visible only when the resonance condition $\omega\approx \Delta$ is reached near $T_c$.  Indeed, one could expect that at least for moderate disorder the main effect of impurities is to enhance the overall coupling of the Higgs mode to light, preserving its frequency structure with the same broad resonance at $2\Delta$ found in the clean case\cite{schwarz_prb20}. In this view, even if subleading the Higgs mode could manifest whenever the matching condition with the pump frequency or with a coupled resonance makes it overcome the overall BCS signal. If confirmed, such a view could finally open the way to a detection mechanism able to disentangle the Higgs response from the BCS quasiparticle one, and to investigate its role on the long-sought unconventional pairing mechanism at play in cuprate superconductors.

\section{Acknowledgements}
We acknowledge financial support by ERC under the project MORE-TEM ERC-SYN (grant agreement No 951215), and by
Sapienza University of Rome under projects Ateneo 2021 (RM12117A4A7FD11B) and Ateneo 2022 (RP1221816662A977). The work of G.S. is supported by the Deutsche Forschungsgemeinschaft under SE 806/20-1.

\appendix\label{appa}
\section{Power series expansion for the density matrix} \label{sec:thg}
Here we outline a 'real space scheme' to evaluate the deviation of the density
matrix from its equilibrium result when the system is perturbed by the coupling
to a time dependent vector potential.
This approach is based on the
equation of motion of the density matrix which allows us for the evaluation of
linear and non-linear current responses for large lattices. We restrict the 
formalism to the BCS limit, i.e. keeping the
local densities and order parameter values constant upon perturbing the system.
In principle,  the approach can be generalized
to include also the collective modes, i.e. the fluctuations of the SC order-parameter amplitude and phase, and the density\cite{seibold_prb21,udina_faraday22}.  However, this
would significantly increases the dimension of the resulting matrix equations, making it numerically too challenging for the present $d$-wave SC model.

Denoting the density matrix with
\begin{equation}
  {\cal R}=\left(
  \begin{array}{cc}
    \rho & \kappa^\dagger \\
    \kappa & \bar{\rho}
  \end{array}\right),
\end{equation}
the equation
of motion reads\cite{ripka}
\begin{equation}\label{eq:eqmot}
  i \frac{d}{dt}{\cal R} =\left\lbrack {\cal R}
  ,{\cal H}^{BdG}\right\rbrack,
\end{equation}
where the BdG hamiltonian matrix
can be derived from the BdG mean field energy as
\begin{equation}
  {\cal H}_{ij}=\frac{\partial E^{BdG}}{\partial {\cal R}_{ji}} \,
\end{equation}
and the latter is given by
\begin{eqnarray*}
  E^{BdG}=-\sum_{ij}t_{ij}\left(\rho_{ij}-\bar{\rho}_{ij}\right)
  -\frac{J}{4} \sum_{\langle ij\rangle} \kappa^\dagger_{ij}\kappa_{ij} \\
  +\sum_i V_i\left( \rho_{ii}-\bar{\rho}_{ii}+1\right) \,.
\end{eqnarray*}  
Here we denote with $\rho$ and $\kappa$ the normal and anomalous averages,
\begin{eqnarray*}
  \rho_{ij}&=&\langle c_{i,\uparrow}^\dagger c_{j,\uparrow}\rangle \\
  \bar{\rho}_{ij}&=&\langle c_{i,\downarrow} c_{j,\downarrow}^\dagger\rangle \\
  \kappa^\dagger_{ij} &=& \langle c_{i,\uparrow}^\dagger c_{j,\downarrow}^\dagger\rangle \\
\kappa_{ij} &=&  \langle c_{i,\downarrow} c_{j,\uparrow}\rangle \,.
\end{eqnarray*}
These are represented in the inhomogeneous systems by $L\times L$ matrices, with the lattice size $L$ varying from 52 to 68 in our simulations.

In the absence of an external field the density matrix ${\cal R}$ and the Hamiltonian ${\cal H}^{BdG}$ commute, so it simply follows from Eq.\ (\ref{eq:eqmot}) that  the density matrix has no time evolution. The dynamics of ${\cal R}(t)$ is induced via the coupling to the electromagnetic field $\vec{E}(t)=-\partial \vec{A}(t)/\partial t $.  
 Let us consider e.g the case of a (spatially constant) field along the $x$ direction. $A_x(t)$ is coupled to the system via the Peierls
substitution $c_{i+x,\sigma}^\dagger c_{i,\sigma} \rightarrow
e^{i A_x(t)} c_{i+x,\sigma}^\dagger c_{i,\sigma}$, where for simplicity we will drop form the equations all the constant by putting the lattice spacing, the electronic charge $e$, the light velocity $c$ and the Planck constant $\hbar$ equal to one. As one can easily check, the interaction terms in Eq.\ (\ref{eq:model}) are unaffected by the Peierls substitution, while the kinetic-energy part gets modified,  leading to the following contribution to $E^{BdG}$:
\begin{eqnarray}
  T^{BdG}&=&-t\left\lbrace e^{iA_x}\rho_{i+x,i}+e^{-iA_x}\rho_{i-x,i}\right. \nonumber\\ &-& \left. e^{-iA_x}\bar{\rho}_{i+x,i}-e^{iA_x}\bar{\rho}_{i+x,i}\right\rbrace \nonumber \\
  &-&t'\left\lbrace e^{iA_x}\rho_{i+x,i+y}+e^{-iA_x}\rho_{i-x-y,i}\right. \nonumber\\ &-& \left. -e^{-iA_x}\bar{\rho}_{i+x+y,i}-e^{iA_x}\bar{\rho}_{i+x+y,i}\right. \nonumber\\
  &+&\left. e^{iA_x}\rho_{i+x,i-y}+e^{-iA_x}\rho_{i-x+y,i}\right. \nonumber\\
  &-&\left. e^{-iA_x}\bar{\rho}_{i+x-y,i}-e^{iA_x}\bar{\rho}_{i+x-y,i}\right\rbrace
  \,.\label{eq:kina}
\end{eqnarray}
Writing $A_x(t)=A_0 f(t)$ one can now expand the density matrix and the hamiltonian (i.e. the kinetic part resulting from Eq. (\ref{eq:kina})) in powers of $A_0$, i.e.
\begin{equation}\label{eq:expand}
  {\cal R}=\sum_{n=0}A_0^n {\cal R}^{(n)}
\end{equation}
where ${\cal R}^{(0)}$ is the equilibrium density matrix
for which
\begin{equation}
  \left\lbrack {\cal R}^{(0)}
  ,{\cal H}^{BdG}\right\rbrack=0 \,,
\end{equation}
as we already emphasized above. 
The current density is obtained at all orders in $A_x$ as $j_x=-(1/N)\frac{\partial E^{BdG}}{\partial A_x}= -(1/N)\frac{\partial T^{BdG}}{\partial A_x}$ with
$N\equiv L^2$.
Thus, by retaining terms up
to third order in $A_x$ one has:
\begin{equation}\label{eq:jxx}
  j_x=\left(1-\frac{1}{2}A_x^2\right) j_{para}^x + A_x\left(1-\frac{1}{6}A_x^2\right) j_{dia}^x
\end{equation}
with
\begin{eqnarray*}
  j_{para}^x&=& it\sum_n\left\lbrack \rho_{n+x,n}-\bar{\rho}_{n-x,n}
  -\rho_{n-x,n}+\bar{\rho}_{n+x,n}\right\rbrack \\
  &+& it'\sum_n\left\lbrack \rho_{n+x+y,n}-\bar{\rho}_{n-x-y,n}
  -\rho_{n-x-y,n}+\bar{\rho}_{n+x+y,n}\right\rbrack \\
  &+& it'\sum_n\left\lbrack \rho_{n+x-y,n}-\bar{\rho}_{n-x+y,n}
  -\rho_{n-x+y,n}+\bar{\rho}_{n+x-y,n}\right\rbrack \\
  j_{dia}^x&=& -t\sum_n\left\lbrack \rho_{n+x,n}-\bar{\rho}_{n-x,n}
  +\rho_{n-x,n}-\bar{\rho}_{n+x,n}\right\rbrack \\
  &-& t'\sum_n\left\lbrack \rho_{n+x+y,n}-\bar{\rho}_{n-x-y,n}
  +\rho_{n-x-y,n}-\bar{\rho}_{n+x+y,n}\right\rbrack \\
  &-& t'\sum_n\left\lbrack \rho_{n+x-y,n}-\bar{\rho}_{n-x+y,n}
  +\rho_{n-x+y,n}-\bar{\rho}_{n+x-y,n}\right\rbrack \,.
\end{eqnarray*}
Here the subscript $para$ and $dia$ refer to the usual identification of the leading terms coupling the gauge field to the fermionic operators in the Hamiltonian, i.e. the linear coupling between the paramagnetic term and $A_x$, and a quadratic coupling between the electronic density and $A_x^2$, that leads to the standard diamagnetic contribution to the current in linear response. To make a closer connection to standard notation in Nambu operators for the SC state
\cite{aoki_prb15,cea_prb16,cea_prb18,silaev_prb19,shimano_prb19,tsuji_prr20,schwarz_prb20,fiore_prb22}, the paramagnetic term is described by a $\tau_0$ Pauli matrix and the diamagnetic term by a $\tau_3$ matrix. Here the $j_{para}^x$ and $j_{dia}^x$ terms represent directly average values of such fermionic operators in the presence of the external gauge field, and such are expressed in terms of the density-matrix elements $\rho$, which in turn contain the dependence on $A_x$ at all orders.  
The expansion Eq. (\ref{eq:expand}) therefore explicitly reads 
\begin{eqnarray*}
  j_{para}^x&=& \sum_n A_0^n j_{para}^{x,(n)} \\
  j_{dia}^x&=& \sum_n A_0^n j_{dia}^{x,(n)}
\end{eqnarray*}
which upon inserting into Eq. (\ref{eq:jxx}) allows us to extract the various
current contributions to order $n$,  $j_x^{(n)}$.

In particular, the 3rd harmonic contribution
to the current density reads
\begin{equation}\label{eq:jxxx}
  j_x^{(3)}=j_{para}^{x,(3)}-\frac{1}{2}A_0^2 j_{para}^{x,(1)}
  + A_0 j_{dia}^{x,(2)}-\frac{1}{6}A_0^3 j_{dia}^{x,(0)}
\end{equation}
where we find that, similar to the isotropic s-wave case \cite{seibold_prb21},
the dominant para- and diamagnetic contributions
are given by $j_{para}^{x,(3)}$ and $A_0 j_{dia}^{x,(2)}$.
On the other hand, $j_{para}^{x,(1)}$ and $j_{dia}^{x,(0)}$ also
enter the calculation of the optical conductivity in first order,
cf. next subsection.

\subsection{First order}\label{sec:1st}
The first order current contribution, relevant for the evaluation
of the optical conductivity, is given by
\begin{equation}
  j_x^{(1)}=j_{para}^{x,(1)}+A_0 j_{dia}^{x,(0)}
\end{equation}
which requires evaluation of the density matrix up to
order $n=1$. 

By selecting all terms $\sim A_0$ in the equation of motion
Eq. (\ref{eq:eqmot}) one obtains
\begin{equation}\label{eq:a1}
  i \underline{\underline{\dot{R}}}^{(1)} =\left\lbrack
  \underline{\underline{R}}^{(1)}
  ,\underline{\underline{H}}^{(0)}\right\rbrack+f(t)
\left\lbrack
  \underline{\underline{R}}^{(0)}
  ,\underline{\underline{V}}\right\rbrack  
\end{equation}
with 
\begin{equation}\label{eq:V}
  \underline{\underline{V}}=\left(
  \begin{array}{cc}
    \underline{\underline{v}} & \underline{\underline{0}} \\
      \underline{\underline{0}} & \underline{\underline{v}}
  \end{array} \right)
\end{equation}
and
\begin{eqnarray}
  v_{nm}&=& -it\left\lbrack \delta_{m,n+x}-\delta_{m,n-x}\right\rbrack \nonumber\\
  &-& it' \left\lbrack \delta_{m,n+x+y}-\delta_{m,n-x-y}\right\rbrack \nonumber \\
  &-& it' \left\lbrack \delta_{m,n+x-y}-\delta_{m,n-x+y}\right\rbrack
\end{eqnarray}

The non-perturbed hamiltonian $\underline{\underline{H}}^{(0)}$
(i.e. for $A_0=0$) can be diagonalized
\begin{eqnarray*}
  \underline{\underline{\tilde{H}}}^{(0)}&=&\underline{\underline{T}}^{-1}
  \underline{\underline{H}}^{(0)}\underline{\underline{T}} \\
  &=&\left(
  \begin{array}{cccccc}
    -E_N & \dots & 0 &0 & \dots& 0\\
    \vdots & \ddots & 0 & \vdots   &\dots  & \vdots\\
    0 & \dots & -E_1     & 0 &\dots & 0 \\
    0 & \dots & 0 & E_1 & \dots &0 \\
    \vdots&  & \vdots & \vdots & \ddots & \vdots\\
    0 &\dots &0 &0 &\dots & E_N
  \end{array}\right)
\end{eqnarray*}
and the same transformation also diagonalizes the non-perturbed
density matrix
\begin{eqnarray*}
  \underline{\underline{\tilde{R}}}^{(0)}&=&\underline{\underline{T}}^{-1}
  \underline{\underline{R}}^{(0)}\underline{\underline{T}} \\
  &=&\left(
  \begin{array}{cccccc}
    1 & \dots & 0 &0 & \dots& 0\\
    \vdots & \ddots & 0 & \vdots   &\dots  & \vdots\\
    0 & \dots & 1     & 0 &\dots & 0 \\
    0 & \dots & 0 & 0 & \dots &0 \\
    \vdots&  & \vdots & \vdots & \ddots & \vdots\\
    0 &\dots &0 &0 &\dots & 0
  \end{array}\right) \,.
\end{eqnarray*}
With this transformation Eq. (\ref{eq:a1}) can be written as
\begin{equation}
  i \dot{\tilde{R}}^{(1)}_{nm}=(E_{mm}-E_{nn}) \tilde{R}^{1}_{nm} +
  (\tilde{R}^{0}_{nn}-\tilde{R}^{0}_{mm})\tilde{V}_{nm} f(t)
 \end{equation}
where $\underline{\underline{\tilde{V}}}$ denotes the transformed matrix
Eq. (\ref{eq:V}).

Consider now a harmonic vector potential $f(t)=\cos{\Omega t}$. Then the
Fourier transformed correction
\begin{equation}
  \tilde{R}^{1}_{nm}(\omega)=\int dt e^{i\omega t} \tilde{R}^{1}_{nm}(t)
\end{equation}
is given by
\begin{eqnarray}
  \tilde{R}^{1}_{nm}(\omega)&=&  \frac{\tilde{R}^{0}_{nn}-\tilde{R}^{0}_{mm}}{\omega-E_{mm}+E_{nn}} \tilde{V}_{nm} \pi\left\lbrack \delta(\omega-\Omega)+\delta(\omega+\Omega\right\rbrack \nonumber\\
  &\equiv & \tilde{\chi}_{nm}(\omega)\tilde{V}_{nm} \pi \left\lbrack \delta(\omega-\Omega)+\delta(\omega+\Omega\right\rbrack \label{eq:r1}
\end{eqnarray}
which can be transformed back to yield the first order
perturbation $R_{ij}^{(1)}$ to the density matrix in the original
site representation. For the presentation of the results in Sec.
\ref{resultsigma}, $\tilde{\chi}_{nm}(\omega\to \omega-i\eta)$
has been computed with a small
imaginary part in order to avoid singularities.

\subsection{Second order}
We proceed by evaluating the diamagnetic contribution to the
third harmonic current $A_0 j_{dia}^{x,(2)}$, cf. Eq. (\ref{eq:jxxx}).
Collecting all term $\sim A_0^2$ we find for the correction to the density
matrix in second order
\begin{eqnarray}
  i \underline{\underline{\dot{R}}}^{(2)}(t) &=&\left\lbrack
  \underline{\underline{R}}^{(2)}(t)
  ,\underline{\underline{H}}^{(0)}\right\rbrack
+\left\lbrack
  \underline{\underline{R}}^{(1)}(t)
  ,\underline{\underline{V}}\right\rbrack f(t) \nonumber \\
&+& \frac{1}{2}\left\lbrack
  \underline{\underline{R}}^{(0)}
  ,\underline{\underline{C}}\right\rbrack f^2(t) \label{eq:a2}
\end{eqnarray}
where we have defined the matrix
\begin{equation}\label{eq:C}
  \underline{\underline{C}}=\left(
  \begin{array}{cc}
    \underline{\underline{c}} & \underline{\underline{0}} \\
      \underline{\underline{0}} & -\underline{\underline{c}}
  \end{array} \right)
\end{equation}
and
\begin{eqnarray}
  c_{nm}&=& t\left\lbrack \delta_{m,n+x}+\delta_{m,n-x}\right\rbrack \nonumber\\
  &+& t' \left\lbrack \delta_{m,n+x+y}+\delta_{m,n-x-y}\right\rbrack \nonumber \\
  &+& t' \left\lbrack \delta_{m,n+x-y}+\delta_{m,n-x+y}\right\rbrack \,.
\end{eqnarray}
Fourier transformation yields
\begin{eqnarray}
  \omega \underline{\underline{{R}}}^{(2)}(\omega) &=&\left\lbrack
  \underline{\underline{R}}^{(2)}(\omega)
  ,\underline{\underline{H}}^{(0)}\right\rbrack \nonumber \\
&+&\frac{1}{2}\left\lbrack
  \underline{\underline{R}}^{(1)}(\omega+\Omega)+\underline{\underline{R}}^{(1)}(\omega-\Omega)
  ,\underline{\underline{V}}\right\rbrack \nonumber \\
&+& \left\lbrack
  \underline{\underline{R}}^{(0)}
  ,\underline{\underline{C}}\right\rbrack \frac{\pi}{2}
  \left\lbrace \delta(\omega) \right.\nonumber \\
  &+&\left. \frac{1}{2}\left\lbrack
  \delta(\omega+2\Omega)+\delta(\omega-2\Omega)\right\rbrack\right\rbrace\label{eq:a22}
\end{eqnarray}
which upon inserting Eq. (\ref{eq:r1}) and diagonalizing can be solved for the
second order contribution to the density matrix as
\begin{eqnarray}
  \tilde{R}^{2}_{nm}(\omega)&=& \frac{\pi}{2}\tilde{\chi}_{nm}(\omega) \tilde{C}_{nm}\left\lbrace \delta(\omega)\right. \nonumber \\
  &+& \left.\frac{1}{2}\left\lbrack \delta(\omega+2\Omega)+\delta(\Omega-2\Omega)\right\rbrack\right\rbrace \nonumber \\
  &+&\frac{\pi}{2}\frac{1}{\omega-E_{mm}+E_{nn}}\left\lbrack \delta(\omega+
  2\Omega)+\delta(\omega)\right\rbrack
 \nonumber \\&\times&  \left\lbrack \underline{\underline{\tilde{\chi}(\omega+\Omega)\tilde{V}}}, \underline{\underline{\tilde{V}}}\right\rbrack_{nm} \nonumber \\
  &+&\frac{\pi}{2}\frac{1}{\omega-E_{mm}+E_{nn}}\left\lbrack \delta(\omega-
  2\Omega)+\delta(\omega)\right\rbrack \nonumber \\
  &\times&  \left\lbrack \underline{\underline{\tilde{\chi}(\omega-\Omega)\tilde{V}}}, \underline{\underline{\tilde{V}}}\right\rbrack_{nm}
    \label{eq:r2}
\end{eqnarray} 
and $\left\lbrack \underline{\underline{\tilde{\chi}(\omega)\tilde{V}}}\right\rbrack_{nm}$ has to understood as $\tilde{\chi}_{nm}(\omega)\tilde{V}_{nm}$.

\subsection{Third order}
Finally, we evaluate the paramagnetic contribution to the
third harmonic current $j_{para}^{x,(3)}$.
Collecting all terms $\sim A_0^3$ in the equation of motion
Eq. (\ref{eq:eqmot}) results in the following equation
for the third order correction to the density
matrix
\begin{eqnarray}
  i \underline{\underline{\dot{R}}}^{(3)}(t) &=&\left\lbrack
  \underline{\underline{R}}^{(3)}(t)
  ,\underline{\underline{H}}^{(0)}\right\rbrack
+\left\lbrack
  \underline{\underline{R}}^{(2)}(t)
  ,\underline{\underline{V}}\right\rbrack f(t) \nonumber \\  
&+& \frac{1}{2}\left\lbrack
  \underline{\underline{R}}^{(1)}
  ,\underline{\underline{C}}\right\rbrack f^2(t) \nonumber \\
&-& \frac{1}{6} \left\lbrack
  \underline{\underline{R}}^{(0)}(t)
  ,\underline{\underline{V}}\right\rbrack f^3(t) \,.
  \label{eq:a3}
\end{eqnarray}

The solution for the contribution at $\omega=3\Omega$ is then
given by
\begin{eqnarray}
  \tilde{R}^{3}_{nm}(3\Omega)&=& -\frac{\pi}{24}\tilde{\chi}_{nm}(3\Omega) \tilde{V}_{nm}     \label{eq:r33}\\
  &+&\frac{\pi}{8}\frac{1}{3\Omega-E_{mm}+E_{nn}}\left\lbrack
\underline{\underline{\tilde{\chi}(2\Omega)}} \underline{\underline{\tilde{C}}},\underline{\underline{\tilde{V}}}  \right\rbrack_{nm} \nonumber\\
  &+&\frac{\pi}{8}\frac{1}{3\Omega-E_{mm}+E_{nn}}\left\lbrack
\underline{\underline{\tilde{\chi}(\Omega)}} \underline{\underline{\tilde{V}}},\underline{\underline{\tilde{C}}}  \right\rbrack_{nm} \nonumber\\
&+&\frac{\pi}{4}\frac{1}{3\Omega-E_{mm}+E_{nn}} \nonumber \\
&\times&\left\lbrack
\frac{1}{2\Omega-E_{mm}+E_{nn}}\left\lbrack
\underline{\underline{\tilde{\chi}(2\Omega)}} \underline{\underline{\tilde{V}}},\underline{\underline{\tilde{V}}}  \right\rbrack , \underline{\underline{\tilde{V}}} \right\rbrack_{nm} \nonumber \,.
\end{eqnarray} 
Not that in case of an isotropic s-wave SC with minimum energies
$E_n=\pm \Delta$, Eq. (\ref{eq:r33}) has contributions at
frequencies $3\Omega=2\Delta$ and $2\Omega=2\Delta$ in agreement
with Ref. \cite{silaev_prb19}.

\section{Averaging over finite size effects}\label{appb}
\begin{figure}[hhh]
\includegraphics[width=8cm,clip=true]{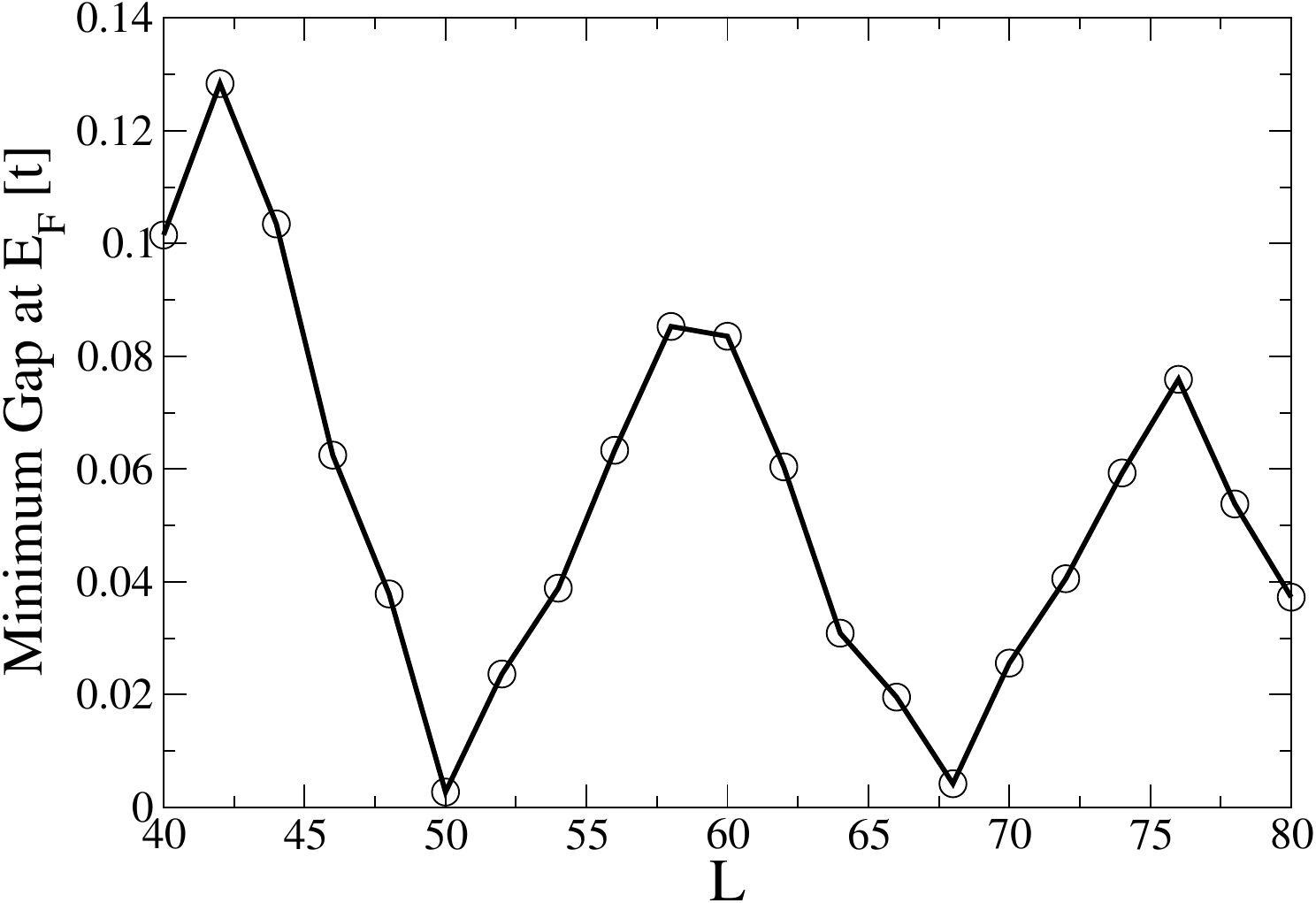}
\caption{The minimum SC gap in a homogeneous d-wave superconductor
  on a $L \times L$ lattice. 
$J/t=1$, $n=0.875$, $t'/t=-0.2$.}     
\label{figapb1}                                                   
\end{figure}

On finite lattices the SC gap is strongly influenced by the
number of $k$-points, in particular for d-wave systems where
the gap vanishes along the nodal direction. This is shown in Fig.\ \ref{figapb1}, where the minimum energy $E_{min}$ is shown for a homogeneous $L\times L$ system and and parameters used in the main paper, with $L$ ranging  from 40 to 80. As one can see, $E_{min}$ reveals an 'oscillatory' behavior of  as a function of $L$, due to the fact that the minimum
gap depends on the 'closeness' of a $k$-point to the intersection
of the underlying Fermi surface with the zone diagonal.  In all cases the maximum spectral gap is $\approx 0.6t$. For a
$60\times 60$ lattice one finds
a minimum SC gap  of $\approx 0.08t$ while for a $68\times 68$ lattice
one has $k$-points closer to the intersection between the Fermi surface and the
zone diagonal, so that the minimum gap $E_{min}$ is smaller ($\approx 0.004t$).

\begin{figure}[hhh]
\includegraphics[width=8cm,clip=true]{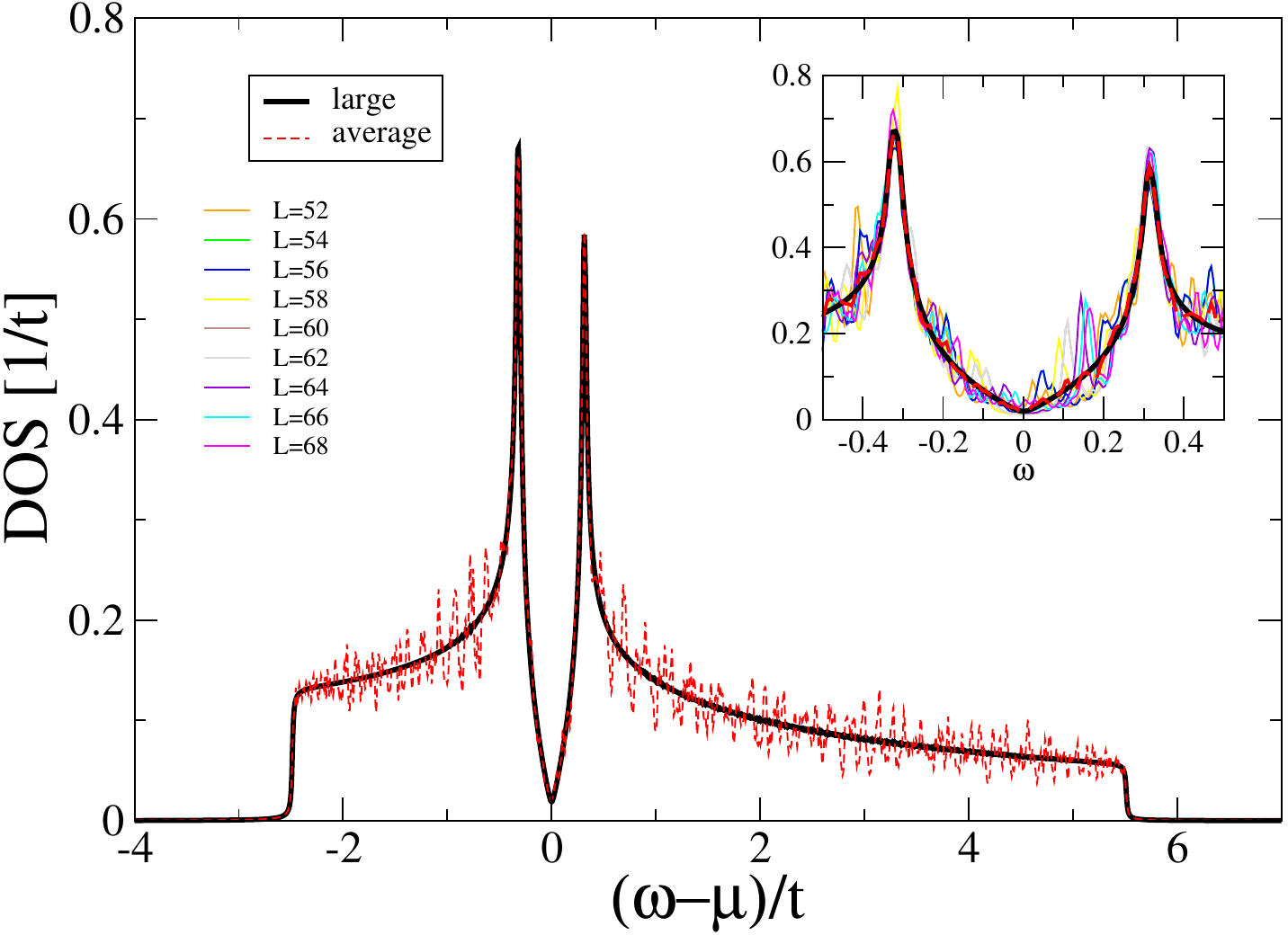}
\caption{Main panel: Density of states of a d-wave superconductor on
  a $800 \times 800$ lattice (black) compared to the average of spectral
  functions on $L \times L$ lattices with $52 \le L\le 68$ (red dashed).
  Inset: DOS from the individual finite lattices in the gap region.
$J/t=1$, $n=0.875$, $t'/t=-0.2$.}     
\label{figapb2}                                                   
\end{figure}  
Despite such an oscillatory behavior, for the lattice sites shown in Fig.\ \ref{figapb1}  the set of $k$ points which is sampled 
within such a period constitutes a mesh in momentum space dense enough to be representative of the behavior on a much larger lattice. This is demonstrated in Fig. \ref{figapb2}. Here we show the comparison between the DOS of a large ($L=800$) system and the DOS obtained as average of the one computed over one period of oscillations for $E_{min}$ in Fig. \ref{figapb1}. More specifically, the average red curve is obtained by averaging the single DOS obtained for $L$ ranging from 52 to 68, that are also shown for comparison in the inset. Here one sees that while the individual densities still retain large oscillations for energies within the maximum SC gap, the averaged DOS is very close to the one obtained for the large $L=800$ lattice, while some residual finite-size oscillations only survive for energies larger than
  the maximum gap. To minimize the effects of the finite lattice size we employed the same approach also for the computation of the results in the inhomogeneous case, performing not only an average over disorder but also an average over different lattice sites.

  \section{Universal conductivity of d-wave superconductors}\label{sec:s0}
  Within a Boltzmann-type approach, the DC conductivity of a disordered
  superconductor can be written as \cite{graf_prb96}
  \begin{equation}\label{eq:si}
    \sigma_{xx}=e^2 N_F v_F^2 {\cal T}_{xx}
  \end{equation}
  where $N_F$ is the normal-state DOS at the Fermi energy and ${\cal T}_{xx}$ plays the role of an effective transport time, that is defined by the equation
  \begin{equation}\label{eq:tau}
    {\cal T}_{xx}=\frac{\gamma^2}{v_F^2}\left\langle \frac{v_{F,x}^2(\phi)}{\sqrt{\Delta^2(\phi)+\gamma^2}^3}\right\rangle \,.
  \end{equation}
  Here $\langle\dots\rangle$ denotes the average over the Fermi surface, $v_F^2\equiv \langle v_{F,x}^2(\phi)\rangle$, and $\gamma$ is the width of the impurity band determined from
  \begin{equation}
    1=\Gamma_{imp}\frac{\left \langle\frac{1}{\sqrt{\gamma^2+\Delta^2(\phi)}}\right\rangle}{C^2+\left\langle\frac{\gamma}{\sqrt{\gamma^2+\Delta^2(\phi)}}\right\rangle^2} \,.
    \end{equation}
  $\Gamma_{imp}$ is proportional to the impurity concentration and
  $C$ denotes the cotangent of the scattering phase shift.
  We solve Eqs. (\ref{eq:si},\ref{eq:tau}) for the homogeneous model Eq. (\ref{eq:model})
  and parameters defined in Sec. \ref{model} which yield a DOS of $N_F=0.305 1/t$.
  As a result we obtain $\sigma_0=2.38 [e^2/\hbar]$
  in the limit of vanishing disorder (equivalent to $C\rightarrow \infty$ and $\gamma\rightarrow 0$), and $\sigma_0=0.6 [e^2/\hbar]$ in the strong-disorder unitary limit (equivalent to $C=0$).

    
  \section{Determination of the scattering time in the normal state}\label{sec:tau}
In order to extract the scattering time $\tau$ one could in principle fit the
 low-energy numerical normal state optical conductivity to the bare Drude form
  \begin{equation}
    \sigma(\omega)=\frac{\varepsilon_0\omega_p^2}{-i\omega+\frac{1}{\tau}}
  \end{equation}
   where $\omega_p$ denotes the plasma frequency.
However, it is more convenient to start from a    
frequency-dependent optical conductivity represented in
  terms of a memory function $M(\omega)$ \cite{wolfle_prb72}
  \begin{equation}\label{eq:sigw}
    \sigma(\omega)=\frac{i\varepsilon_0 \omega_p^2}{\omega+M(\omega)}
  \end{equation}
 which allows us to fit the numerical
  result over a larger frequency range and therefore yields a more robust value
  for $\tau$.

  \begin{figure}[htb]
\includegraphics[width=8cm,clip=true]{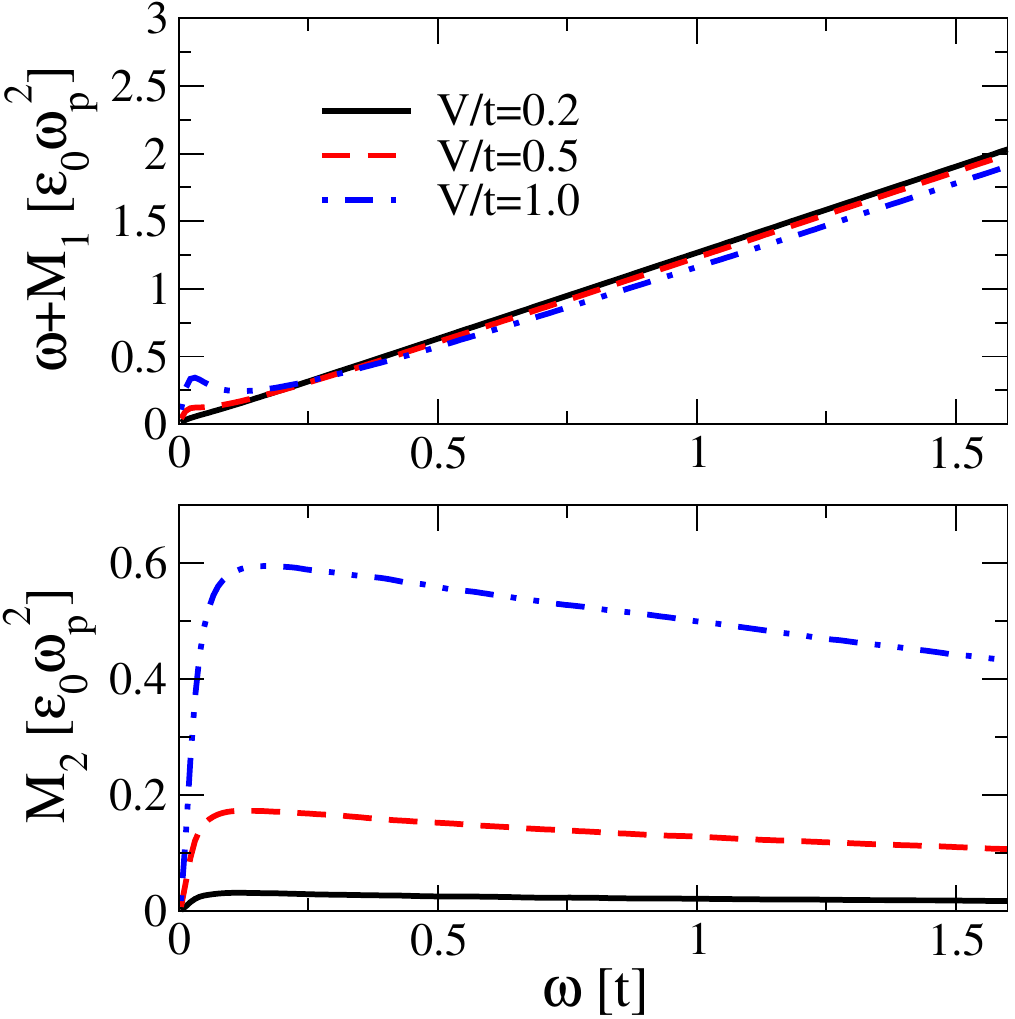}
\caption{Memory function for different disorder levels
  as extracted from the conductivities. 
  $J/t=1$, $n=0.875$, $t'/t=-0.2$.}     
\label{figmem}                                                   
\end{figure}  
  
\begin{figure}[htb]
\includegraphics[width=8cm,clip=true]{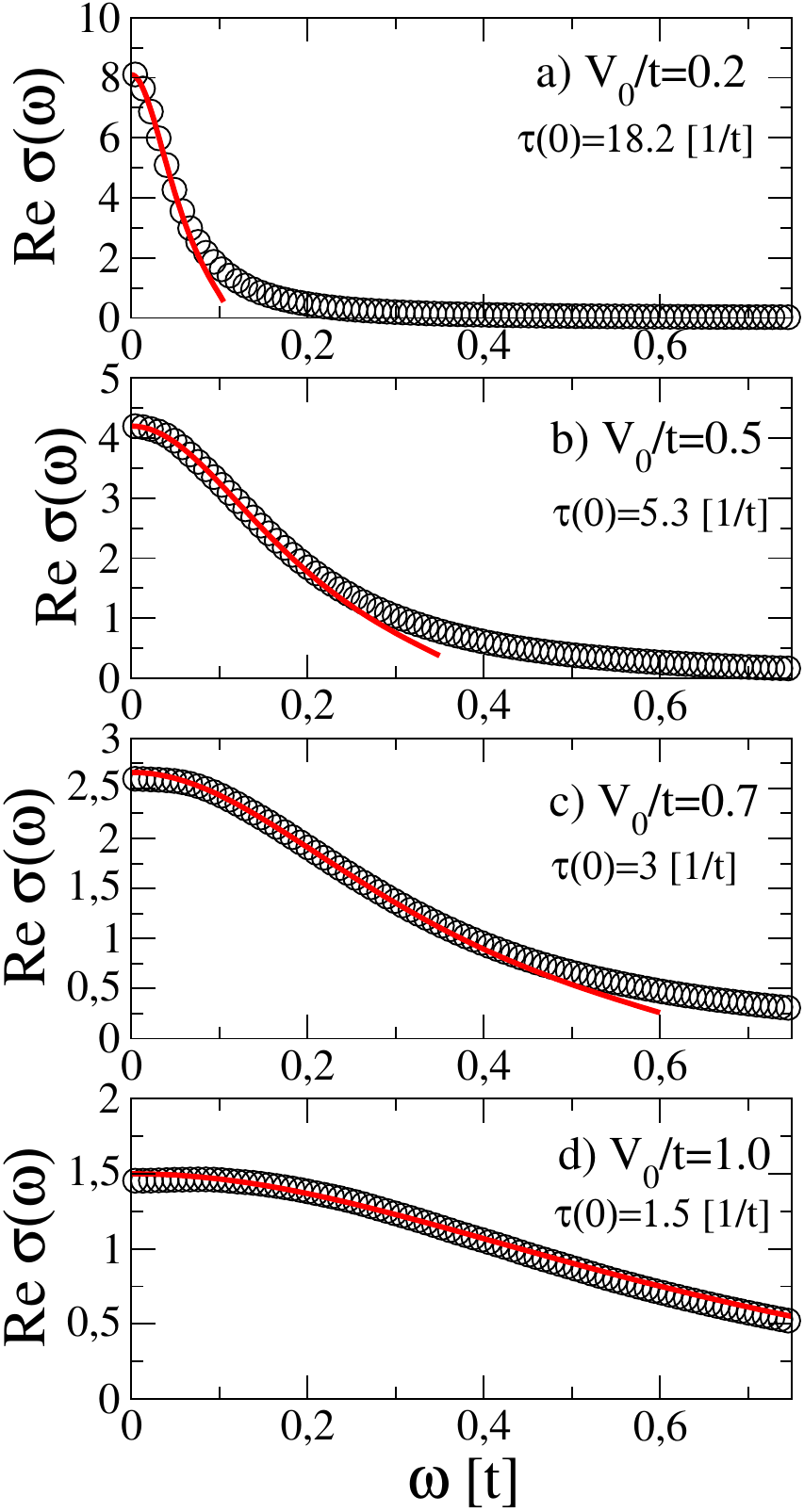}
\caption{The real part of the normal state conductivity (circles) at different
  disorder levels together
  with the fit from
  Eq. (\ref{eq:sigfit}) with the memory function Eq. (\ref{eq:mem}).
  $J/t=1$, $n=0.875$, $t'/t=-0.2$.}     
\label{figtau}                                                   
\end{figure}

  Eq. (\ref{eq:sigw}) can be rewritten in a Drude-like form
  \begin{equation}
    \sigma(\omega)=\varepsilon_0\omega_p^2\frac{g(\omega)}{-i\omega+\frac{1}{\tau(\omega)}}
  \end{equation}
  with
  \begin{eqnarray*}
    g(\omega)&=&\frac{1}{1+\frac{M_1(\omega)}{\omega}} \\
    \frac{1}{\tau(\omega)}&=& g(\omega)M_2(\omega)
  \end{eqnarray*}
so that the real part $\sigma_1(\omega)$ can be
  obtained from
  \begin{equation}\label{eq:sigfit}
    \sigma_1(\omega)=\varepsilon_0\omega_p^2\frac{g(\omega)\tau(\omega)}{1+\omega^2\tau^2(\omega)} \,.
    \end{equation}

Fig. \ref{figmem} shows the real $M_1$ and imaginary $M_2$ part of the memory
function for three different disorder levels. We can approximately fit $M(\omega)$ with the formula
  \begin{equation}\label{eq:mem}
    \omega+M(\omega) \approx A \ln\frac{B+\omega}{B-\omega}+i A \Theta(B-|\omega|) \,.
  \end{equation}
  This approximation reproduces the linear behavior of $M_1(\omega)\sim \omega$
  for small frequencies (approximately below $\omega/t\simeq 0.5$) and replaces the imaginary part $M_2(\omega)$ by
  a constant value above a threshold scale $B$. Such an approximations can be used to extrapolate $M_2$ at small frequencies in the regime where finite-size uncertainties set in. From Eq.\ (\ref{eq:mem}) the 
 frequency-dependent scattering time reads:
  \begin{equation}
    \tau(\omega)=\frac{1}{\omega}\ln\frac{B+\omega}{B-\omega}
  \end{equation}
  with the zero frequency limit $\tau\equiv \tau(0)=2/B$.
  Fig. \ref{figtau} shows that Eq. (\ref{eq:mem}) provides a good
  fits to the low-frequency conductivity. The corresponding zero-frequency scattering times derived from the above formula at each disorder level  are reported in the panels.

We briefly comment on an alternative possibility, namely determining
the scattering rate from the imaginary part $\sigma_2(\omega)$ in the normal
state, cf. red dashed curves in the right panels of Fig. \ref{occup}.
From a conventional Drude picture one would expect
\begin{displaymath}
  \omega\sigma_2(\omega)=\frac{\sigma_0}{\tau}
  \frac{\omega^2\tau^2}{1+\omega^2\tau^2}\,
\end{displaymath}
so that at $\omega\tau=1$ the value of $\omega\sigma_2$ should be
half the value at $\omega\to\infty$. Estimating the relaxation time
from this approach yields slightly larger values than from the fitting
of $\sigma_1(\omega)$ within the scheme outlined above,
e.g. $\tau=4/t$ instead of $\tau=3/t$ for $V/t=0.7$. In fact, fitting
the memory function results a more accurate
description of the conductivity  at low energies so that we consider
these values of $\tau$ as more reliable.

\bibliography{bibl}





\end{document}